\documentclass[pre,twocolumn,amsmath,amssymb,floatfix,superscriptaddress,nopacs]{revtex4}

\usepackage{graphicx}
\usepackage{latexsym}
\usepackage{amsmath}
\usepackage{amssymb}
\usepackage{amsfonts}
\usepackage{bm}
\usepackage{color}

\newcommand{\Fcal}{\ensuremath{{\cal F}}}

\newcommand{\Lcal}{\ensuremath{{\cal L}}}

\newcommand{\Ocal}{\ensuremath{{\cal O}}}

\newcommand{\ddiff}{\ensuremath{\mathrm{d}}}

\newcommand{\bfzero}{\ensuremath{{\bf 0}}}

\newcommand{\evec}{\mbox{$\bf e$}}
\newcommand{\hvec}{\mbox{$\bf h$}}
\newcommand{\mvec}{\mbox{$\bf m$}}

\newcommand{\pvec}{\mbox{$\bf p$}}
\newcommand{\qvec}{\mbox{$\bf q$}}
\newcommand{\rvec}{\mbox{$\bf r$}}

\newcommand{\svec}{\mbox{$\bf s$}}
\newcommand{\tvec}{\mbox{$\bf t$}}
\newcommand{\uvec}{\mbox{$\bf u$}}
\newcommand{\vvec}{\mbox{$\bf v$}}
\newcommand{\Avec}{\mbox{$\bf A$}}
\newcommand{\Bvec}{\mbox{$\bf B$}}
\newcommand{\Cvec}{\mbox{$\bf C$}}

\newcommand{\Tvec}{\mbox{$\bf T$}}
\newcommand{\qhat}{\ensuremath{{\hat{q}}}}

\newcommand{\qhatvec}{\ensuremath{{\hat{\bf q}}}}
\newcommand{\rhatvec}{\ensuremath{{\hat{\bf r}}}}
\newcommand{\thetar}{\ensuremath{\theta_r}}
\newcommand{\thetaq}{\ensuremath{\theta_q}}

\newcommand{\Tgen}{\ast}
\newcommand{\Trot}{\prime}
\newcommand{\Tref}{\neg}

\newcommand{\la}{\left<}
\newcommand{\ra}{\right>}
\newcommand{\kB}{k_\mathrm{B}}

\newcommand{\agrid}{a_{\mathrm{grid}}}
\newcommand{\nLgrid}{n_L}
\newcommand{\nVgrid}{n_V}
\newcommand{\tincr}{\delta \tau}
\newcommand{\tsamp}{\Delta \tau}

\newcommand{\taualph}{\tau_{\alpha}}
\newcommand{\taustar}{\tau_{\star}}
\newcommand{\tauexp}{\tau_{\mathrm{exp}}}
\newcommand{\tauA}{\tau_{\mathrm{A}}}
\newcommand{\taubasin}{\tau_{\mathrm{b}}}

\newcommand{\Nt}{\ensuremath{N_\mathrm{t}}}
\newcommand{\Nc}{\ensuremath{N_\mathrm{c}}}

\newcommand{\sigab}{\sigma_{\alpha\beta}}

\newcommand{\Sab}{\bar{\sigma}_{\alpha\beta}}

\newcommand{\Sxy}{\bar{\sigma}_{12}}

\newcommand{\sabQ}{\sigma^{\mathrm{q}}_{\alpha\beta}}

\newcommand{\cone}{c_\mathrm{L}}
\newcommand{\ctwo}{c_\mathrm{N}}
\newcommand{\cthree}{c_{\perp}}
\newcommand{\cfour}{c_\mathrm{T}}

\newcommand{\Cabcd}{\bar{c}_{\alpha\beta\gamma\delta}}
\newcommand{\Cxxxx}{\bar{c}_{1111}}
\newcommand{\Cyyyy}{\bar{c}_{2222}}

\newcommand{\Cxxyy}{\bar{c}_{1122}}

\newcommand{\Cxyxy}{\bar{c}_{1212}}
\newcommand{\Cxxxy}{\bar{c}_{1112}}
\newcommand{\Cyyyx}{\bar{c}_{2221}}
\newcommand{\Cabcdprime}{\bar{c}^{\prime}_{\alpha\beta\gamma\delta}}
\newcommand{\Cxxxxprime}{\bar{c}^{\prime}_{1111}}
\newcommand{\Cyyyyprime}{\bar{c}^{\prime}_{2222}}
\newcommand{\Cxxyyprime}{\bar{c}^{\prime}_{1122}}
\newcommand{\Cxyxyprime}{\bar{c}^{\prime}_{1212}}
\newcommand{\Cone}{\bar{c}_\mathrm{L}}
\newcommand{\Ctwo}{\bar{c}_\mathrm{N}}
\newcommand{\Cthree}{\bar{c}_{\perp}}
\newcommand{\Cfour}{\bar{c}_\mathrm{T}}

\newcommand{\Etilde}{e}

\newcommand{\ISFab}{\bar{\sigma}^{\circ}_{\alpha\beta}}

\newcommand{\ISFlongi}{\bar{\sigma}^{\circ}_{11}}
\newcommand{\ISFshear}{\bar{\sigma}^{\circ}_{12}}
\newcommand{\ISFnorm}{\bar{\sigma}^{\circ}_{22}}
\newcommand{\ISFnormQ}{\bar{\sigma}^{\circ\mathrm{q}}_{22}}

\newcommand{\ISFnormS}{\sigma^{\circ\mathrm{s}}_{22}}
\newcommand{\Tglass}{T_{\mathrm{g}}}

\newcommand{\tauL}{\tau_{\mathrm{L}}}
\newcommand{\tauT}{\tau_{\mathrm{T}}}
\newcommand{\pmax}{p_{\mathrm{max}}}
\newcommand{\qmin}{q_{\mathrm{min}}}

\newcommand{\scut}{s_\mathrm{cut}}
\newcommand{\smin}{s_\mathrm{min}}
\newcommand{\sighatab}{\hat{\sigma}_{\alpha\beta}}
\newcommand{\sighatidab}{\hat{\sigma}^{\mathrm{id}}_{\alpha\beta}}
\newcommand{\sighatexab}{\hat{\sigma}^{\mathrm{ex}}_{\alpha\beta}}

\newcommand{\Pid}{P_{\mathrm{id}}}
\newcommand{\Pex}{P_{\mathrm{ex}}}

\newcommand{\Eabcd}{\ensuremath{E_{\alpha\beta\gamma\delta}}}
\newcommand{\Jabcd}{\ensuremath{J_{\alpha\beta\gamma\delta}}}
\newcommand{\epsab}{\epsilon_{\alpha\beta}}

\newcommand{\cabcd}{\ensuremath{c_{\alpha\beta\gamma\delta}}}

\newcommand{\cinf}{c_{\infty}}

\begin{document}

\title{Correlations of tensor field components in isotropic systems\\
with an application to stress correlations in elastic bodies}

\author{J.P.~Wittmer}
\affiliation{Institut Charles Sadron, Universit\'e de Strasbourg \& CNRS, 23 rue du Loess, 67034 Strasbourg Cedex, France}
\author{A.N. Semenov}
\affiliation{Institut Charles Sadron, Universit\'e de Strasbourg \& CNRS, 23 rue du Loess, 67034 Strasbourg Cedex, France}
\author{J. Baschnagel}
\affiliation{Institut Charles Sadron, Universit\'e de Strasbourg \& CNRS, 23 rue du Loess, 67034 Strasbourg Cedex, France}

\date{\today}

\begin{abstract}
Correlation functions of components of second-order tensor fields in isotropic systems can be reduced 
to an isotropic forth-order tensor field characterized by a few invariant correlation functions (ICFs). 
It is emphasized that components of this field depend in general on the coordinates of 
the field vector variable and thus on the orientation of the coordinate system.  
These angular dependencies are distinct from those of ordinary anisotropic systems. 
As a simple example of the procedure to obtain the ICFs we discuss correlations of time-averaged stresses 
in isotropic glasses where only one ICF in reciprocal space becomes a finite constant $\Etilde$
for large sampling times and small wavevectors. 
It is shown that $\Etilde$ is set by the typical size of the frozen-in stress components normal 
to the wavevectors, i.e. it is caused by the symmetry breaking of the stress for each independent configuration.
Using the presented general mathematical formalism for isotropic tensor fields
this finding explains in turn the observed long-range stress correlations in real space.
Under additional but rather general assumptions $\Etilde$ is shown to be given by a thermodynamic quantity,
the equilibrium Young modulus $E$. We thus relate for certain isotropic amorphous bodies the existence of finite 
Young or shear moduli to the symmetry breaking of a stress component in reciprocal space.
\end{abstract}
\maketitle

\section{Introduction}
\label{intro}


\paragraph*{Tensorial foundation of science and engineering.}
The fundamental laws of physics and the constitutive relations of engineering
are formulated in terms of {\em tensors} and {\em tensor fields}
(assigning a tensor to each point of the mathematical space)
\cite{McConnell,Schultz_Piszachich,Lambourne,TadmorCMTBook} 
which by construction guarantees them to hold independently of the
specific coordinate system used for their description. 
Moreover, also instabilities and failure in materials science and engineering, 
e.g., for granular piles and silos \cite{NeddermanBook} or 
plastic deformation in soft or glassy materials \cite{Argon79,Langer98,Bocquet04,Voigtmann14,Fielding14}, 
must be described by appropriate {\em tensorial invariants} of tensor fields
and this should also be crucial in principle for mesoscopic computer models \cite{Tanguy11,Barrat18}
of localized plastic failure of a broad range of systems.
 
\paragraph*{Isotropic systems.}
Isotropic systems, such as generic isotropic elastic bodies \cite{LandauElasticity,TadmorCMTBook}, 
simple and complex fluids \cite{HansenBook,RubinsteinBook}, 
amorphous metals and glasses \cite{DonthGlasses}, polymer foams and networks \cite{RubinsteinBook}
or, as a matter of fact, our entire universe \cite{Lambourne} are described (at least on some scales)
by {\em isotropic tensors} and {\em isotropic tensor fields}
\cite{McConnell,TadmorCMTBook,Schultz_Piszachich}.
Let us assume for simplicity that the system is not only isotropic but also
spatially homogeneous \cite{foot_homogeneous}, achiral \cite{LandauElasticity}, 
stationary in time and embedded in a $d$-dimensional Euclidean vector space 
described by an orthonormal Cartesian tensor basis \cite{TadmorCMTBook}.
A point in this vector space is either called $\rvec$ (real space) or $\qvec$ (reciprocal space).
It is well known that the components of isotropic tensors remain unchanged under an
orthogonal coordinate transformation \cite{TadmorCMTBook}.  
For instance, the component $E_{1212}$ of the forth-order elastic modulus tensor 
$\Eabcd$ of an isotropic body \cite{LandauElasticity,TadmorCMTBook} 
is always given by the shear modulus $\mu$, i.e. an invariant material property.
(See Sec.~\ref{ten_ten} and Sec.~\ref{e2E_viscoelast} for details.)
Interestingly, this does in general not hold for the components of isotropic tensor {\em fields}
\cite{Schultz_Piszachich} which may depend (in real space)
not only on the length $r$ of the field vector $\rvec$ 
but also on the (normalized) coordinate dependent 
components $\hat{r}_{\alpha}$ of its direction $\hat{\rvec} = \rvec/r$.
This implies that mathematically and physically legitimate isotropic tensor field 
components may depend on the orientation of the coordinate system and this, as we shall see, 
holds in a related manner both in real and in reciprocal space. 
As we shall emphasize these angular dependencies differ from those of 
ordinary anisotropic systems with {\em frame-invariant} angular-dependent material functions,
say for crystalline solids \cite{LandauElasticity}.


\begin{figure}[t]
\centerline{\resizebox{0.95\columnwidth}{!}{\includegraphics*{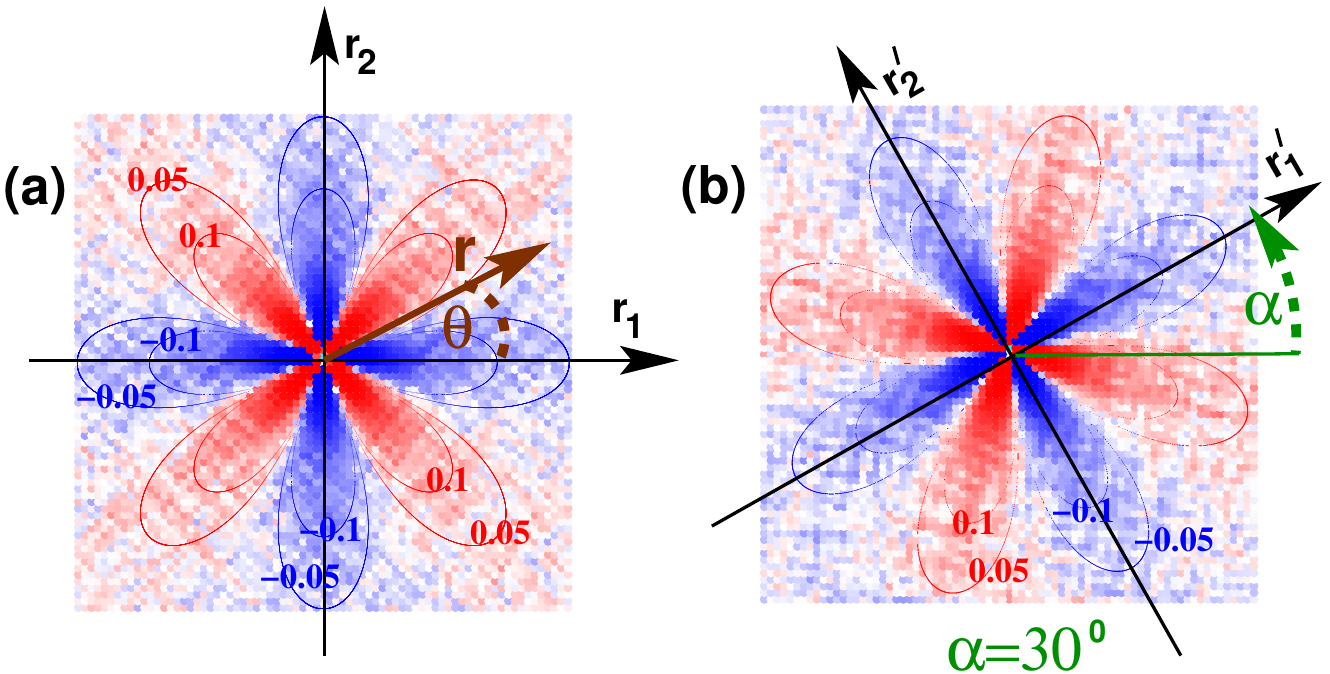}}}
\caption{Autocorrelation function $\Cxyxy(\rvec)$ of time-averaged shear-stress fields
$\Sxy(\rvec)$ of a two-dimensional isotropic elastic body:
{\bf (a)} Unrotated frame with coordinates $(r_1,r_2)$,
{\bf (b)} frame ($r_1^{\prime},r_2^{\prime})$ rotated by an angle 
$\alpha=30^{\circ}$ (rotations marked by ``$\prime$").
Albeit the system is isotropic, the correlation function is strongly angle dependent revealing 
an octupolar symmetry and depends, moreover, on the orientation of the coordinate system.
While each pixel corresponds in {\bf (a)} and {\bf (b)} to the same spatial position $\rvec$,
the correlation functions differ by the angle $\alpha$.
$\Cxyxy(\rvec)$ is negative (blue) along the axes
and positive (red) along the bisection lines of the respective axes.
The theoretical prediction Eq.~(\ref{eq_Cxyxy_intro}) is indicated by thin lines.
}
\label{fig_intro}
\end{figure}

\paragraph*{Invariant correlation functions.} 
Tensor fields are probed experimentally or in computer simulations
by means of correlation functions of some of their components.
For instance, as shown in Fig.~\ref{fig_intro} for an isotropic elastic body discussed in more detail below, 
one may investigate the spatial correlations of the shear-stress component of the {\em stress} tensor field
\cite{Bocquet04,Fuchs17,Fuchs18,Fuchs19,lyuda18,Lemaitre15,Lemaitre18,Harrowell16,lyuda22a,MM22}.
For isotropic systems such correlation functions must be isotropic tensor fields.
They may thus depend on the orientation of the coordinate systems 
as demonstrated by the example given in panel (b) of Fig.~\ref{fig_intro}.
Another example of current interest are the correlations of {\em strain} tensor field components of a broad 
range of isotropic systems which have also been shown to reveal ``anisotropic" correlation functions
\cite{Bocquet04,Bocquet09,Barrat14c,FlennerSzamel15b,Weeks15,Fuchs18b,Reichman21,Reichman21b}.
Importantly, correlation functions of tensor components describe the linear response
due to a small perturbation \cite{HansenBook}, say an inclusion in an elastic body 
\cite{Eshelby57,Eshelby59}.
Such a tensorial response field
may thus be angle-dependent for systems and source terms which are both perfectly isotropic.
(See Sec.~\ref{ten_pointsource} for a comparison of tensorial response fields
and associated correlation fields in isotropic systems.)
As we shall remind \cite{Lemaitre15}
and emphasize in this study, while the isotropy of the system may not be manifested 
by {\em one} correlation function of tensor field components it is nevertheless present in the 
general mathematical structure of the {\em complete set} of {\em all} 
correlation functions of the investigated tensor field.
While this makes the interpretation of observed tensor field pattern,
like the ones given in Fig.~\ref{fig_intro}, more intricate, the good new is that
this complete set of correlation functions is determined by a small number of 
``Invariant Correlation Functions" (ICFs) \cite{Lemaitre15,Fuchs18,lyuda18}.
We emphasize here that theory and computational studies should focus on
these ICFs and this in a first step in reciprocal space (cf. Appendix~\ref{FT}).  
The real-space correlations are then obtained by inverse Fourier transformation (FT). 


\paragraph*{Specific case considered.}
We demonstrate the general procedure by an analysis of stress field correlations in 
two-dimensional ($d=2$) simulated isotropic elastic bodies \cite{LandauElasticity,TadmorCMTBook}. 
While previous work characterizes correlations of the instantaneous
stress field \cite{Fuchs17,Fuchs18,Fuchs19,lyuda18,Harrowell16,lyuda22a,MM22}
or of the stress field of the system's ``inherent states" (local energy minima)
\cite{Lemaitre15,Lemaitre18},
we rather compute the time-averaged stress fields $\Sab(\qvec)$ 
in reciprocal space ($\qvec$ being the wavevector) for each 
independent configuration $c$ and analyze their correlation functions $\Cabcd(\qvec)$.
This is best done using ``Natural Rotated Coordinates" (NRC) aligned with $\qvec$ allowing 
the precise determination of the ICFs characterizing the isotropic tensor field. 
Importantly, only one of these ICFs is shown in the hydrodynamic limit and for 
sufficiently large sampling times $\tsamp$ to become a finite constant $\Etilde >0$.
This phenomenological constant characterizes the typical size of the frozen stress components normal 
to each wavevector $\qvec$, i.e. it measures the (continuous) symmetry breaking of the stress field 
in reciprocal space for each independent configuration.
This finding and a proper treatment of tensor field correlations reflecting 
the material isotropy directly imply that the stress correlations in real space must be long-ranged.
For instance, the correlation function of the shear-stress field must decay as
\begin{equation}
\Cxyxyprime(\rvec) \simeq - \frac{\Etilde}{4\pi r^2}\cos[4(\theta-\alpha)]
\label{eq_Cxyxy_intro}
\end{equation}
for sufficiently large $r = |\rvec|$ with 
$\theta$ being the angle of the field vector $\rvec$ in the unrotated physical system,
Fig.~\ref{fig_intro}(a), and $\alpha$ the rotation angle of the coordinate system, Fig.~\ref{fig_intro}(b). 
(Rotations of the coordinate system are marked by primes ``$\prime$".)
We thus confirm recent computer simulations on stress correlations in binary Lennard-Jones glasses
\cite{Lemaitre14,Lemaitre15,Lemaitre17,Lemaitre18,lyuda22a}
and more general theoretical considerations \cite{Fuchs17,Fuchs18,Fuchs19,lyuda18} 
on supercooled liquids and amorphous elastic bodies.
   
\paragraph*{Outline.}
We begin by reviewing in Sec.~\ref{ten} general features of tensor fields relevant 
for isotropic and achiral systems \cite{McConnell,Schouten,Schultz_Piszachich,TadmorCMTBook}.
The main computational points (model system, data production)
are summarized in Sec.~\ref{algo} before we turn in Sec.~\ref{res} to our 
central numerical results. 
Taking advantage of recent theoretical studies \cite{lyuda18,lyuda22a,Fuchs17,Fuchs18,Fuchs19}
we explain in Sec.~\ref{e2E} why the only phenomenological parameter $\Etilde$ needed to fit our data 
should be similar to a thermodynamic quantity, the equilibrium Young modulus $E$.
A summary of the presented work and an outlook are given in Sec.~\ref{conc}.
Appendix~\ref{FT} summarizes some properties of FTs \cite{abramowitz,numrec}
while Appendix~\ref{q2r} addresses the FT of the relevant correlation functions in $d=2$.
More details on our simulation model and on the computation of the local stress fields
are given in Appendix~\ref{comp} and Appendix~\ref{stress}.
%

\section{Review of isotropic tensor fields}
\label{ten}

\subsection{Introduction}
\label{ten_intro}

\paragraph*{Generalities.}
Familiarity with the general ideas and notations of tensor algebra and analysis,
as developed systematically in the standard textbooks
\cite{McConnell,Schouten,Schultz_Piszachich,TadmorCMTBook}, 
is taken for granted.
We remind that a tensor field assigns a tensor to each point of the mathematical space,
in our case a $d$-dimensional Euclidean vector space \cite{Schultz_Piszachich}.  
An element of this vector space is denoted by the ``spatial position" $\rvec$ in real space
or by the ``wavevector" $\qvec$ for the corresponding Fourier transformed reciprocal space.
The relations for tensor fields are formulated below in reciprocal space since this is more 
convenient both on theoretical and numerical grounds due to the assumed spatial
translational invariance. The properties of the corresponding 
real space tensor field are then obtained by inverse FT.

\paragraph*{Tensor components and basis.}
We assume for simplicity Cartesian coordinates with an orthonormal basis 
$\{\evec_1,\ldots,\evec_d\}$ \cite{McConnell,Schultz_Piszachich,TadmorCMTBook}.
Greek letters $\alpha, \beta,\ldots$ are used for the indices of the tensor (field) components. 
A twice repeated index $\alpha$ is summed over the values $1,\ldots,d$, e.g., 
$\qvec = q_{\alpha} \evec_{\alpha}$ with $q_{\alpha}$ standing for the vector coordinates. 
This work is chiefly concerned with tensors
\begin{equation}
\Tvec^{(o)} = T_{\alpha_1\ldots\alpha_o} \evec_{\alpha_1} \ldots \evec_{\alpha_o}
\label{eq_ten_intro_Trank_def}
\end{equation}
of ``order" (or ``rank") $o=2$ and $o=4$ and their corresponding tensor fields 
with components depending either on $\rvec$ or $\qvec$. 
As common we refer to a tensor (field) $\Tvec^{(o)}$
by indicating its components $T_{\alpha_1\ldots\alpha_o}$. 
The order of a component is given by the number of suffixes.
Note that
\begin{equation}
T_{\alpha_1\ldots\alpha_o}(\qvec) = \Fcal[T_{\alpha_1\ldots\alpha_o}(\rvec)]
\label{eq_ten_intro_FT}
\end{equation}
for the $d^o$ components in real and reciprocal space.

\paragraph*{Transforms.}
We consider linear orthogonal coordinate transformations (marked by ``$\Tgen$")
$\evec_{\alpha}^{\Tgen} = c_{\alpha\beta} \evec_{\beta}$ with matrix coefficients
$c_{\alpha\beta}$ given by the direction cosine 
$c_{\alpha\beta} \equiv \cos(\evec_{\alpha}^{\Tgen},\evec_{\beta})$ \cite{Schultz_Piszachich}.
$c_{\alpha\beta} = \delta_{\alpha\beta}$ if nothing is changed.
For a simple reflection of, say, the $1$-axis and a rotation in the $12$-plane
by an angle $\theta$ we have, respectively,
\begin{eqnarray}
\mbox{reflection} &:& c_{1\beta}=-\delta_{1\beta}
\label{eq_ten_intro_reflection} \\
\mbox{rotation} &:&
\begin{array}[t]{ll}
c_{11}=c_{22}  & = \cos(\theta) \\
c_{12}=-c_{21} & = \sin(\theta)
\end{array}
\label{eq_ten_intro_rotation}
\end{eqnarray}
and $c_{\alpha\beta}=\delta_{\alpha\beta}$ for all other indices.
We remind that \cite{Schultz_Piszachich}
\begin{equation}
T^{\Tgen}_{\alpha_1\ldots\alpha_o}(\qvec) =
c_{\alpha_1\nu_1} \ldots c_{\alpha_o\nu_o} T_{\nu_1\ldots\nu_o}(\qvec)
\label{eq_ten_intro_orthtrans}
\end{equation}
under a general orthogonal transform.
For a reflection of the $1$-axis we thus have, e.g.,
\begin{eqnarray}
T_{12}^{\Tgen}(\qvec) &=& -T_{12}(\qvec), T_{11}^{\Tgen}(\qvec) = T_{11}(\qvec),
\label{eq_ten_intro_refl1axis} \\
T^{\Tgen}_{1222}(\qvec) & = & -T_{1222}(\qvec),
T^{\Tgen}_{1221}(\qvec) = T_{1221}(\qvec),
\nonumber
\end{eqnarray}
i.e. quite generally we have sign inversion for an {\em odd number} 
of indices $\alpha,\beta,\ldots$ equal to the index of the inverted axis.
Please note that the field vector 
$\qvec=q_{\alpha}\evec_{\alpha}=q_{\alpha}^{\Tgen}\evec_{\alpha}^{\Tgen}$ 
remains unchanged by these ``passive" transforms albeit its coordinates change.

\paragraph*{Outline.}
We come back to this issue in the next subsection, Sec.~\ref{ten_iso}.
The symmetries of tensors and tensor fields relevant for the present work are summarized
in Sec.~\ref{ten_sym}.
Isotropic tensors are presented in Sec.~\ref{ten_ten},
general isotropic tensor fields of order $1 \le o \le 4$ in Sec.~\ref{ten_field} and,
more specifically,
forth-order isotropic tensor fields $T_{\alpha\beta\gamma\delta}(\qvec)$ for $d=2$
in Sec.~\ref{ten_Tabcd_d2}.
Finally, Sec.~\ref{ten_pointsource} outlines some general relations 
for second-order tensor fields $R_{\alpha\beta}(\rvec)$ corresponding to the ``linear response" 
caused by a point-like ``source term" $S_{\alpha\beta}(\rvec) = s_{\alpha\beta} \delta(\rvec)$
in real space.

\subsection{Isotropic tensors and tensor fields}
\label{ten_iso}

\paragraph*{Isotropic tensors.}
Isotropic systems are described by ``isotropic tensors" and ``isotropic tensor fields".
Components of an isotropic tensor remain unchanged by {\em any} orthogonal coordinate transformation 
\cite{Schultz_Piszachich,TadmorCMTBook}, i.e.
\begin{equation}
T_{\alpha_1\ldots\alpha_o}^{\Tgen} =T_{\alpha_1\ldots\alpha_o}.
\label{eq_ten_iso}
\end{equation}
As noted at the end of Sec.~\ref{ten_intro} the sign of tensor components change 
for a reflection of one axis 
if the number of indices equal to the inverted axis is {\em odd}.
Consistency with Eq.~(\ref{eq_ten_iso}) implies that 
{\em all tensor components with an odd number of equal indices must vanish}, e.g., 
\begin{equation}
T_{12}=T_{1112} = T_{1222} = T_{1234} = T_{1344} = 0.
\label{eq_ten_iso_oddnumber_vanish}
\end{equation}

\paragraph*{Isotropic tensor fields.}
The corresponding isotropy condition for tensor fields is given by \cite{Schultz_Piszachich}
\begin{equation}
T_{\alpha_1\ldots\alpha_o}^{\Tgen}(q_1,\ldots,q_d) 
= T_{\alpha_1\ldots\alpha_o}(q_1^{\Tgen},\ldots,q_d^{\Tgen}) 
\label{eq_ten_iso_field}
\end{equation}
which reduces to Eq.~(\ref{eq_ten_iso}) for $\qvec=\bfzero$.
Please note that the fields on the left hand side of Eq.~(\ref{eq_ten_iso_field})
are evaluated with the original coordinates while the fields on the right hand side
are evaluated with the transformed coordinates.
Another way to state this is to say that the left hand fields are computed at
the original vector $\qvec=(q_1,\ldots,q_d)$ while the right hand fields are computed at
the ``actively transformed" vector $\qvec^{\Tgen} = (q_1^{\Tgen},\ldots,q_d^{\Tgen})$.
It is for this reason that Eq.~(\ref{eq_ten_iso_oddnumber_vanish})
does not hold in general for tensor fields,
i.e. finite components with an odd number of equal indices, e.g., $T_{1222}(\qvec) \ne 0$,
are possible in principle for finite wavevectors.

\paragraph*{Natural Rotated Coordinates.}
Fortunately, there are convenient coordinates,
called ``Natural Rotated Coordinates" (NRC), where the nice symmetry 
Eq.~(\ref{eq_ten_iso_oddnumber_vanish}) for isotropic tensors can be also used
for tensor fields. Let us assume that the wavevector $\qvec$ points into 
the direction of one of the axes, say, $\qvec = q \delta_{1\beta}$ with $q = |\qvec|$.
(This may be achieved by a first rotation of the coordinate system.)
Let us denote by ``$\Tref$" arbitrary inversions of axes in this frame.
We may thus rewrite Eq.~(\ref{eq_ten_iso_field}) as 
$T_{\alpha_1\ldots\alpha_o}^{\Tref}(q_1) = T_{\alpha_1\ldots\alpha_o}(q_1^{\Tref})$
since $q_2=q_2^{\Tref}=\ldots=q_d=q_d^{\Tref}=0$.
If we now assume in addition that $T_{\alpha_1\ldots\alpha_o}(\qvec)$
is an {\em even} function of its field variable $\qvec$ this becomes
\begin{equation}
T_{\alpha_1\ldots\alpha_o}^{\Tref}(q) = T_{\alpha_1\ldots\alpha_o}(q),
\label{eq_ten_iso_fieldNCF}
\end{equation}
i.e. both sides only depend on the same scalar parameter.
We may thus use for each $q$ the same reasoning as for tensor components.

\paragraph*{Product theorem for isotropic tensor fields.}
Let us state a useful theorem for a general tensor field 
$\Cvec(\qvec) = \Avec(\qvec) \otimes \Bvec(\qvec)$ 
with $\Avec(\qvec)$ and $\Bvec(\qvec)$ being two isotropic tensor fields
and $\otimes$ standing either for an outer product, 
e.g. $C_{\alpha\beta\gamma\delta}(\qvec)=A_{\alpha\beta}(\qvec)B_{\gamma\delta}(\qvec)$,
or an inner product, 
e.g. $C_{\alpha\beta\gamma\delta}(\qvec)=A_{\alpha\beta\gamma\nu}(\qvec)B_{\nu\delta}(\qvec)$.
Hence,
\begin{eqnarray}
\Cvec^{\Tgen}(\qvec) & = & \left(\Avec(\qvec) \otimes \Bvec(\qvec)\right)^{\Tgen} 
= \Avec^{\Tgen}(\qvec) \otimes \Bvec^{\Tgen}(\qvec) \nonumber \\
& = & \Avec(\qvec^{\Tgen}) \otimes \Bvec(\qvec^{\Tgen}) = \Cvec(\qvec^{\Tgen})
\label{eq_ten_iso_AB2C}
\end{eqnarray}
using in the second step a general property of tensor (field) products, due to 
Eq.~(\ref{eq_ten_intro_orthtrans}), and in the third step
Eq.~(\ref{eq_ten_iso_field}) for the fields $\Avec(\qvec)$ and $\Bvec(\qvec)$
where $\qvec^{\Tgen}$ stands for the ``actively" transformed field position.
We have thus demonstrated that $\Cvec(\qvec)$ is also an isotropic tensor field.
This theorem allows to construct isotropic tensor fields from known isotropic 
tensor fields $\Avec(\qvec)$ and $\Bvec(\qvec)$. 
\paragraph*{Multilinear forms.}
Alternatively, isotropic tensor fields may be constructed using
multilinear forms  \cite{Schultz_Piszachich,TadmorCMTBook}
\begin{equation}
L(\vvec^1,\ldots,\vvec^o;\qvec) = T_{\alpha_1\ldots\alpha_o}(\qvec) \
v^1_{\alpha_1}\ldots v^o_{\alpha_o}
\label{eq_ten_iso_multilinear}
\end{equation}
of order $o$ where $L$ stands for a linear and (first order) homogeneous functional
of a $d$-dimensional vector space \cite{Schultz_Piszachich,TadmorCMTBook}
and $\vvec^1,\ldots,\vvec^o$ are $o$ arbitrary vectors of this vector space
(with superscripts exceptionally used here for the numbering of these vectors).
For tensor fields this functional also depends on the tensor field vector $\qvec$.
The goal is then to construct generic isotropic tensor fields
associated with multilinear forms. This is done in a first step by means of
additive terms of all possible scalars formed with the vectors $\vvec^1,\ldots,\vvec^o$ and $\qvec$,
e.g., inner products $\vvec^1 \cdot \vvec^2$, $\vvec^1 \cdot \qvec$ or $\qvec\cdot \qvec$
or triple products such as $[\vvec^1 \vvec^2 \qvec]$. In a second step all contributions are eliminated
which are yet incompatible with Eq.~(\ref{eq_ten_iso_field}) and other imposed symmetries.
We shall illustrate this below in Sec.~\ref{ten_field}.

\paragraph*{Kronecker and Levi-Civita tensors.}
We note for later convenience that the Kronecker symbol $\delta_{\alpha\beta}$ is 
an invariant tensor, $\delta_{\alpha\beta}^{\Tgen}=\delta_{\alpha\beta}$,
for any orthogonal transform \cite{Schultz_Piszachich}.
As a consequence, any tensor field, only containing additive terms such as 
$i(q) \delta_{\alpha\beta} \delta_{\gamma\delta}$ with $i(q)$ being an invariant scalar,
is an isotropic tensor field. The same applies for tensor fields
with terms containing one or several factors $\qhat_{\alpha}$ since
in agreement with Eq.~(\ref{eq_ten_iso_field}) this implies, e.g,
\begin{equation}
\left( i(q) \qhat_{\alpha} \qhat_{\beta} \delta_{\gamma\delta} \right)^{\Tgen}
 = i(q) \qhat_{\alpha}^{\Tgen} \qhat_{\beta}^{\Tgen} \delta_{\gamma\delta}
\label{eq_ten_iso_qterms} 
\end{equation}
where Eq.~(\ref{eq_ten_intro_orthtrans}) was used.
The situation is more intricate for terms containing the Levi-Civita (``permutation") 
tensor $\varepsilon_{\alpha\beta\gamma}$ \cite{McConnell} 
which is only invariant for the rotation subgroup but 
in general not for reflections \cite{Schultz_Piszachich}. 
%

\subsection{Assumed symmetries}
\label{ten_sym}

All second-order tensors in this work are symmetric, $T_{\alpha\beta}= T_{\beta\alpha}$,
and the same applies for the corresponding tensor fields in either $\rvec$- or $\qvec$-space.
This is, e.g., the case for the stress fields $\sigab(\qvec)$.
We assume for all forth-order tensor fields that
\begin{eqnarray}
T_{\alpha\beta\gamma\delta}(\qvec) & = & T_{\beta\alpha\gamma\delta}(\qvec)=
T_{\alpha\beta\delta\gamma}(\qvec)
\label{eq_ten_sym_A}\\
T_{\alpha\beta\gamma\delta}(\qvec) & = & T_{\gamma\delta\alpha\beta}(\qvec) \mbox{ and } 
\label{eq_ten_sym_B} \\
T_{\alpha\beta\gamma\delta}(\qvec) & = & T_{\alpha\beta\gamma\delta}(-\qvec). \label{eq_ten_sym_C}
\end{eqnarray}
Let us remind that forth-order tensor fields are often constructed by taking
outer products \cite{TadmorCMTBook} of second-order tensor fields.
We consider, e.g., correlation functions
\begin{equation}
T_{\alpha\beta\gamma\delta}(\qvec) = \la \hat{T}_{\alpha\beta}(\qvec) \hat{T}_{\gamma\delta}(-\qvec) \ra
\label{eq_ten_sym_D}
\end{equation}
with $\hat{T}_{\alpha\beta}(\qvec)$ being an instantaneous (not ensemble-averaged) 
second-order tensor field. 
Eq.~(\ref{eq_ten_sym_A}) then follows from the symmetry of the second-order tensor fields.
Evenness, Eq.~(\ref{eq_ten_sym_C}), is a necessary condition for {\em achiral} systems.
It implies that $T_{\alpha\beta\gamma\delta}(\qvec)$ is real if $T_{\alpha\beta\gamma\delta}(\rvec)$ 
is real and, moreover, Eq.~(\ref{eq_ten_sym_B}) for correlation functions since
$\langle \hat{T}_{\alpha\beta}(\qvec) \hat{T}_{\gamma\delta}(-\qvec) \rangle
= \langle \hat{T}_{\gamma\delta}(\qvec) \hat{T}_{\alpha\beta}(-\qvec) \rangle$.
As already emphasized, it is assumed that all our systems are {\em isotropic}.
This implies that Eqs.~(\ref{eq_ten_iso}-\ref{eq_ten_iso_field}) must hold for the 
ensemble-averaged tensor fields.  Since our systems are also {\em achiral}, 
Eq.~(\ref{eq_ten_iso_fieldNCF}) applies and tensor field components with an odd number of 
equal indices must vanish if one axis points into the direction of the wavevector.
We consider in the following subsections isotropic tensors and tensor fields respecting the above symmetries.

\subsection{Isotropic tensors}
\label{ten_ten}

Isotropic tensors of different order are discussed, e.g., in Sec.~2.5.6 of Ref.~\cite{TadmorCMTBook}. 
While all tensors of odd order must vanish, we have
\begin{eqnarray}
T_{\alpha\beta} & = & k_1 \delta_{\alpha\beta}, \label{eq_ten_ten_o2} \\
T_{\alpha\beta\gamma\delta} & = & i_1 \delta_{\alpha\beta} \delta_{\gamma\delta} 
+ i_2 \left(
\delta_{\alpha\gamma} \delta_{\beta\delta} + \delta_{\alpha\delta} \delta_{\beta\gamma}
\right) \label{eq_ten_ten_o4}
\end{eqnarray}
where $k_1$, $i_1$ and $i_2$ are invariant scalars.
Please note that all symmetries stated above hold,
especially also Eq.~(\ref{eq_ten_iso_oddnumber_vanish}).
An example for an isotropic second-order tensor is
the isotropic stress tensor $\sigab = -P \delta_{\alpha\beta}$
(with $P$ being the average normal pressure) which is assumed in the present work.
Note that the symmetry Eq.~(\ref{eq_ten_sym_A}) was used for the second relation, Eq.~(\ref{eq_ten_ten_o4}).
Importantly, this implies that only {\em two} coefficients are needed for a forth-order isotropic tensor.
As further discussed in Sec.~\ref{e2E_viscoelast},
the elastic modulus tensor $\Eabcd$ is thus completely described by
{\em two} elastic moduli, say $\lambda$ and $\mu$, and 
the stress relaxation modulus tensor $\Eabcd(\tau)$ by {\em two} relaxation functions, 
say the ``mixed relaxation function" $M(\tau)$ and 
the ``shear-stress relaxation function" $G(\tau)$ \cite{foot_t_tau}.

\subsection{Tensor fields for isotropic achiral systems}
\label{ten_field}

We begin by summarizing the relevant isotropic tensor fields for $1 \le o \le 4$ 
compatible with the assumed symmetries (cf. Sec.~\ref{ten_sym}). 
With $l_n(q)$, $k_n(q)$, $j_n(q)$  and $i_n(q)$ being invariant scalar functions of $q$ we have
\begin{eqnarray}
T_{\alpha}(\qvec) & = & l_1(q) \ q_{\alpha} 
\label{eq_ten_field_o1} \\
T_{\alpha\beta}(\qvec) & = & 
k_1(q) \ \delta_{\alpha\beta} + k_2(q) \ q_{\alpha}q_{\beta} 
\label{eq_ten_field_o2} \\
T_{\alpha\beta\gamma}(\qvec) & = &
j_1(q) \ q_{\alpha} \delta_{\beta\gamma} 
+ j_2(q) \ q_{\beta}  \delta_{\alpha\gamma} \nonumber \\
& + & j_3(q) \ q_{\gamma} \delta_{\alpha\beta} 
+ j_4(q) \ q_{\alpha} q_{\beta} q_{\gamma}
\label{eq_ten_field_o3} \\
T_{\alpha\beta\gamma\delta}(\qvec) & = &
i_1(q) \ \delta_{\alpha\beta} \delta_{\gamma\delta} \label{eq_ten_field_o4_old} \\
& + & i_2(q) \ \left(
\delta_{\alpha\gamma} \delta_{\beta\delta} + \delta_{\alpha\delta} \delta_{\beta\gamma}
\right) \nonumber \\
& + & i_3(q) \ \left(
q_{\alpha} q_{\beta}\delta_{\gamma\delta} + q_{\gamma}q_{\delta}\delta_{\alpha\beta} 
\right) \nonumber \\
& + & i_4(q) \ q_{\alpha} q_{\beta} q_{\gamma} q_{\delta} \nonumber\\
& + & 
i_5(q)  \left( 
q_{\alpha}q_{\gamma} \delta_{\beta\delta}+  
q_{\alpha}q_{\delta} \delta_{\beta\gamma}+\right. \nonumber \\
& & \hspace*{.90cm}\left. q_{\beta}q_{\gamma} \delta_{\alpha\delta}+  
q_{\beta}q_{\delta} \delta_{\alpha\gamma}  
\right) \nonumber
\end{eqnarray}
with $q_{\alpha} \equiv \qvec \cdot \evec_{\alpha}$.
Let us first check that the stated relations are reasonable.
All relations reduce (continuously) for $\qvec \to \bfzero$ to the isotropic tensors stated in 
Sec.~\ref{ten_ten}; all are, according to the discussion in the last paragraph
of Sec.~\ref{ten_iso}, isotropic tensor fields consistent with Eq.~(\ref{eq_ten_iso_field})
and all symmetries stated in Sec.~\ref{ten_sym} for the second- and forth-order tensor 
fields are satisfied. 
All tensor fields of even (odd) order are even (odd) with respect to $\qvec$.
Hence, tensor fields of odd order vanish for $\qvec \to \bfzero$
consistently with Sec.~\ref{ten_ten}.
Note that the terms due to the invariants $k_1(q)$, $i_1(q)$ and $i_2(q)$
are strictly isotropic and, hence, independent of the coordinate system.
All other terms depend on the components $q_{\alpha}$ 
and thus on the coordinate system. 

Following Refs.~\cite{Schultz_Piszachich,ForsterBook} let us first show that 
Eq.~(\ref{eq_ten_field_o2}) holds. According to Eq.~(\ref{eq_ten_iso_multilinear})
one may represent a general second-order tensor field by a bilinear form $L(\uvec,\vvec;\qvec)$. 
Invariant with respect to orthogonal transformations are the scalars 
$\qvec \cdot \qvec = q^2$, $\qvec \cdot \uvec$, $\qvec \cdot \vvec$
and, additionally, the triple product $[\uvec \vvec \qvec]$ for three-dimensional systems.
We obtain thus the general bilinear form
\begin{eqnarray}
L(\uvec,\vvec;\qvec) & = & k_1 \uvec\cdot \vvec + 
k_2 (\qvec\cdot\uvec) (\qvec\cdot\vvec) 
+ k_3 [\uvec \vvec \qvec]
\nonumber \\
& = & T_{\alpha\beta}(\qvec) u_{\alpha} v_{\beta} \ \mbox{ with } \nonumber \\
T_{\alpha\beta}(\qvec) & = &
k_1 \delta_{\alpha\beta} + k_2 q_{\alpha} q_{\beta}
+ k_3 \varepsilon_{\alpha\beta\gamma} q_{\gamma}
\label{eq_ten_field_o2_B}
\end{eqnarray} 
with $k_1$, $k_2$ and $k_3$ being scalar coefficients.
While the first two terms of $T_{\alpha\beta}(\qvec)$ are fine with respect to Eq.~(\ref{eq_ten_iso_field}),
the last term must be eliminated since $\varepsilon_{\alpha\beta\gamma}$ changes sign
for a reflection at one axis \cite{Schultz_Piszachich}.
 
The indicated relations for the other fields are obtained in a similar manner \cite{Schultz_Piszachich}.
Note that using the product theorem, Eq.~(\ref{eq_ten_iso_AB2C}),
one may obtain the isotropic tensor fields of third and forth order as sums of products
of lower-order isotropic tensor fields. 
For instance, let $A_{\alpha\beta}(\qvec)$ and $B_{\alpha\beta}(\qvec)$
be two second-order isotropic tensor fields according to Eq.~(\ref{eq_ten_field_o2}).
It is readily seen that $A_{\alpha\beta}(\qvec) B_{\gamma\delta}(\qvec) +
A_{\gamma\delta}(\qvec) B_{\alpha\beta}(\qvec)$ immediately
implies the first four terms of Eq.~(\ref{eq_ten_field_o4_old}),
i.e. $i_5(q)=0$ if $T_{\alpha\beta\gamma\delta}(\qvec)$ is {\em only} constructed from 
two isotropic second-order tensors. 
For the indicated more general isotropic forth-order tensors Eq.~(\ref{eq_ten_field_o4_old}) 
with $i_5(q)\ne 0$ we have included contributions due to products 
$T_{\alpha}(\qvec) T_{\beta\gamma\delta}(\qvec)$ of isotropic first- and third-order tensor fields.
 
Let us check that terms containing the Levi-Civita tensor 
$\varepsilon_{\alpha\beta\gamma}$ cannot contribute 
additional terms to a forth-order tensor field
$T_{\alpha\beta\gamma\delta}(\qvec)$ obeying the assumed symmetries.
The forth-order multilinear form 
$L(\uvec,\vvec,\svec,\tvec;\qvec) = T_{\alpha\beta\gamma\delta}(\qvec) 
u_{\alpha} v_{\beta} s_{\gamma} t_{\delta}$
may indeed {\em apriori} contain in $d=3$ terms of products of invariants such as 
\begin{eqnarray}
(\uvec \cdot \vvec) \ \left[\svec \tvec \qvec\right] & = & 
\delta_{\alpha\beta} \varepsilon_{\gamma\delta\nu} q_{\nu}
\ \ \ \ \ \  u_{\alpha} v_{\beta} s_{\gamma} t_{\delta},
\label{eq_ten_field_check1} \\
\left[ \uvec \vvec \qvec \right] \ \left[\svec \tvec \qvec \right] & = & 
\varepsilon_{\alpha\beta\nu} \varepsilon_{\gamma\delta\mu}
q_{\nu} q_{\mu} \ \ u_{\alpha} v_{\beta} s_{\gamma} t_{\delta}.
\label{eq_ten_field_check2}
\end{eqnarray}
Terms of the first type are disallowed for the same reason
as argued for the second-order isotropic tensor, Eq.~(\ref{eq_ten_field_o2_B}).
Note also that such terms would not be compatible with Eq.~(\ref{eq_ten_sym_C}).
Terms of the the second type, Eq.~(\ref{eq_ten_field_check2}),
can be expressed using Sarrus' law as
\begin{eqnarray}
\varepsilon_{\alpha\beta\nu} \varepsilon_{\gamma\delta\mu} q_{\nu} q_{\mu} & = &
\delta_{\alpha\gamma} \delta_{\beta\delta} q^2 
+ \delta_{\alpha\delta} q_{\beta} q_{\gamma} 
+ \delta_{\beta\gamma} q_{\alpha} q_{\delta} 
\nonumber \\
&-&
\delta_{\alpha\delta} \delta_{\beta\gamma} q^2 
- \delta_{\alpha\gamma} q_{\beta} q_{\delta} 
- \delta_{\beta\delta} q_{\alpha} q_{\gamma}. 
\nonumber
\end{eqnarray}
This is, however, in conflict with Eq.~(\ref{eq_ten_sym_A}) and Eq.~(\ref{eq_ten_sym_B}).
To enforce, e.g., the $\alpha \leftrightarrow \beta$-symmetry
the multilinear form must also contain a term  
$\left[ \vvec \uvec \qvec \right] \left[\svec \tvec \qvec \right]$
which is equal to $- \left[ \uvec \vvec \qvec \right] \left[\svec \tvec \qvec \right]$.
All contributions of the second type needed for symmetry reasons thus exactly cancel.

Finally, let us note that for physical reasons it is useful to rewrite for finite wavevectors
($\qvec \ne \bfzero)$ the isotropic forth-order tensor field Eq.~(\ref{eq_ten_field_o4_old}) in terms of the 
components $\qhat_{\alpha} = \qhatvec \cdot \evec_{\alpha}$ of the normalized wavevector $\qhatvec$.
It is thus convenient to bring in factors of $q$ and to redefine
$i_3(q) \to i_3(q)/q^2$, $i_4(q) \to i_4(q)/q^4$ and $i_5(q) \to i_5(q)/q^2$.
We thus rewrite Eq.~(\ref{eq_ten_field_o4_old}) as
\begin{eqnarray}
T_{\alpha\beta\gamma\delta}(\qvec) & = &
i_1(q) \ \delta_{\alpha\beta} \delta_{\gamma\delta} \label{eq_ten_field_o4} \\
& + & i_2(q) \left(
\delta_{\alpha\gamma} \delta_{\beta\delta} + \delta_{\alpha\delta} \delta_{\beta\gamma}
\right) \nonumber \\
& + & i_3(q) \left(
\qhat_{\alpha} \qhat_{\beta}\delta_{\gamma\delta} + \qhat_{\gamma} \qhat_{\delta}\delta_{\alpha\beta} 
\right) \nonumber \\
& + & i_4(q) \ \qhat_{\alpha} \qhat_{\beta} \qhat_{\gamma} \qhat_{\delta} \nonumber\\
& + & 
i_5(q)  \left( 
\qhat_{\alpha} \qhat_{\gamma} \delta_{\beta\delta}+  
\qhat_{\alpha} \qhat_{\delta} \delta_{\beta\gamma}+\right. \nonumber \\
& & \hspace*{.90cm}\left. \qhat_{\beta}\qhat_{\gamma} \delta_{\alpha\delta}+  
\qhat_{\beta}\qhat_{\delta} \delta_{\alpha\gamma}  
\right). \nonumber
\end{eqnarray}
Now all $i_n(q)$ have the same physical units.
As we shall see, the $i_n(q)$ become often constant, $i_n(q) \to i_n$,
or negligibly tiny for sufficiently small (but finite) $q$.

\subsection{Isotropic $T_{\alpha\beta\gamma\delta}(\qvec)$ in two dimensions}
\label{ten_Tabcd_d2}

We have stated in Eq.~(\ref{eq_ten_field_o4}) the general 
form of forth-order tensor fields consistent with the assumed symmetries. 
As shown here, not all indicated terms are needed
for the two-dimensional systems studied numerically in this work. 
To see this let us, following the discussion in Sec.~\ref{ten_iso},
rotate the coordinate system such that the $1$-axis points into the direction of $\qvec$,
i.e. $q_{\alpha}^{\Trot} = q \delta_{1\alpha}$ with the prime ``$\Trot$" marking the rotated frame.
Using this coordinate system we define the four functions \cite{lyuda18}
\begin{equation}
\left.
\begin{array}{ll}
\cone(q)   & \equiv T^{\Trot}_{1111}(\qvec) \\
\ctwo(q)   & \equiv T^{\Trot}_{2222}(\qvec) \\
\cthree(q) & \equiv T^{\Trot}_{1122}(\qvec) \\
\cfour(q)  & \equiv T^{\Trot}_{1212}(\qvec)
\end{array}
\right\} \ \mbox{ for } \ q_{\alpha}^{\Trot} = q \delta_{1\alpha}. 
\label{eq_ten_Tabcd_d2_A} 
\end{equation}
Since the system is isotropic, these functions depend on the wavelength $q$ but not on the direction
$\hat{\qvec}$ of the wavector $\qvec$. In other words, they are {\em invariant} under rotation and 
they do not change either (being only dependent on $q$) if one of the coordinate axes is inverted.
Importantly, all other components $T^{\Trot}_{\alpha\beta\gamma\delta}(\qvec)$ 
are either by Eq.~(\ref{eq_ten_sym_A}) and Eq.~(\ref{eq_ten_sym_B}) identical to these invariants
or must vanish for an odd number of equal indices due to Eq.~(\ref{eq_ten_iso_fieldNCF})
as discussed in Sec.~\ref{ten_iso}.
The $d^o=16$ components $T^{\prime}_{\alpha\beta\gamma\delta}(\qvec)$ are thus completely 
determined by the four invariants and this for any $\qvec$. 
The tensor field $T_{\alpha\beta\gamma\delta}(\qvec)$ in the original frame may then be obtained by the 
inverse rotation of $T^{\prime}_{\alpha\beta\gamma\delta}(\qvec)$ to the original unrotated frame
using Eq.~(\ref{eq_ten_intro_orthtrans}). 
Let us define the coefficients $i_n(q)$ using
\begin{eqnarray}
\cone(q)   & = & i_1(q) + 2i_2(q) + 2i_3(q) + i_4(q) \nonumber \\
\ctwo(q)   & = & i_1(q) + 2i_2(q) \nonumber \\
\cthree(q) & = & i_1(q) + i_3(q) \nonumber \\
\cfour(q)  & = & i_2(q) 
\label{eq_ten_Tabcd_d2_B} 
\end{eqnarray}
which is equivalent to the inverse relations
\begin{eqnarray}
i_1(q) & = & \ctwo(q) - 2\cfour(q) \label{eq_ten_Tabcd_d2_BB} \\
i_2(q) & = & \cfour(q) \nonumber \\
i_3(q) & = & \cthree(q) - \ctwo(q)+ 2\cfour(q)  \nonumber \\
	i_4(q) & = & \cone(q) + \ctwo(q) - 2\cthree(q) - 4\cfour(q). \nonumber
\end{eqnarray}
Consistently with Ref.~\cite{lyuda18} it is then seen that 
\begin{eqnarray}
T_{\alpha\beta\gamma\delta}(\qvec) & = &
i_1(q) \ \delta_{\alpha\beta} \delta_{\gamma\delta} \label{eq_ten_Tabcd_d2_C} \\
& + & i_2(q) \left(
   \delta_{\alpha\gamma} \delta_{\beta\delta} + \delta_{\alpha\delta} \delta_{\beta\gamma}
\right) \nonumber \\
& + & i_3(q) \left(
\qhat_{\alpha}\qhat_{\beta}\delta_{\gamma\delta} + \qhat_{\gamma}\qhat_{\delta}\delta_{\alpha\beta} 
\right) \nonumber \\
& + & i_4(q) \ \qhat_{\alpha} \qhat_{\beta} \qhat_{\gamma} \qhat_{\delta} \nonumber
\end{eqnarray}
which agrees with Eq.~(\ref{eq_ten_field_o4}) if we set $i_5(q) \equiv 0$.
%

\subsection{Response to point sources}
\label{ten_pointsource}

Isotropic tensor fields may also be constructed by taking 
the (inner or outer) product of a tensor field and a (constant) tensor. 
We focus here on the second-order tensor field
\begin{equation}
R_{\alpha\beta}(\qvec) = \frac{1}{V} \ C_{\alpha\beta\gamma\delta}(\qvec) \ s_{\gamma\delta}
\label{eq_ten_ps_A}
\end{equation}
obtained from a forth-order tensor field $C_{\alpha\beta\gamma\delta}(\qvec)$ and a second-order 
tensor $s_{\alpha\beta}$ and where (as always) summation over repeated indices is implied. 
For later convenience we have introduced the volume $V$ of the system.
The results presented below are readily generalized for different types of products
of tensor fields and tensors and for dimensions other than $d=2$.

If both $C_{\alpha\beta\gamma\delta}(\qvec)$ and $s_{\alpha\beta}$ are isotropic, 
the product theorem Eq.~(\ref{eq_ten_iso_AB2C}) discussed in Sec.~\ref{ten_iso} 
implies that $R_{\alpha\beta}(\qvec)$ must also be an isotropic tensor field. 
Under the additional assumptions stated in Sec.~\ref{ten_sym} $R_{\alpha\beta}(\qvec)$ 
is then given by Eq.~(\ref{eq_ten_field_o2}) in terms of two invariants $k_1(q)$ and $k_2(q)$. 
These invariants can in turn be expressed in terms of the invariants of 
$C_{\alpha\beta\gamma\delta}(\qvec)$ and $s_{\alpha\beta}$.
It is important to emphasize that albeit being closely related $R_{\alpha\beta}(\qvec)$ and 
$C_{\alpha\beta\gamma\delta}(\qvec)$ have in general different angular dependences. 
This may be readily seen by focusing on the specific cases $R_{12}(\qvec)$ and $C_{1212}(\qvec)$. 
It follows from Eq.~(\ref{eq_ten_field_o2}) that
$R_{12}(\qvec) \propto \qhat_1 \qhat_2 \propto \sin(2\theta)$ with 
$\qhat_1 = \cos(\theta)$ and $\qhat_2 = \sin(\theta)$
and from Eq.~(\ref{eq_ten_Tabcd_d2_C}) that 
$C_{1212}(\qvec) = i_2(q) + i_4(q) \qhat_1^2 \qhat_2^2$ is given 
by an angular-independent scalar plus a term proportional to $\cos(4\theta)$.
In other words, $R_{12}(\qvec)$ is a quadrupolar field 
whereas $C_{1212}(\qvec)$ is octupolar \cite{foot_multipolar}.
Both fields thus reveal distinct angular patterns.

For reasons which will become obvious below we shall call 
$R_{\alpha\beta}(\qvec)$ the ``response",
$C_{\alpha\beta\gamma\delta}(\qvec)$ the ``correlation function" or ``propagator"
and $s_{\gamma\delta}$ the ``source" or ``perturbation".
Up to now we have not used that the field vector $\qvec$ refers to the
wavevector characterizing fields in reciprocal space.
Using Eq.~(\ref{eq_FT_B}) it is seen that the tensor $s_{\alpha\beta}/V$ in reciprocal space 
corresponds to a ``point source"
$S_{\alpha\beta}(\rvec) = s_{\alpha\beta} \delta(\rvec)$
in real space where we have used Dirac's delta function. The response
$R_{\alpha\beta}(\rvec) = \Fcal^{-1}[R_{\alpha\beta}(\qvec)]$
in real space is then given by 
\begin{equation}
R_{\alpha\beta}(\rvec) = C_{\alpha\beta\gamma\delta}(\rvec) s_{\gamma\delta}
\label{eq_ten_ps_C}
\end{equation}
using the correlation function
$C_{\alpha\beta\gamma\delta}(\rvec) = \Fcal^{-1}[C_{\alpha\beta\gamma\delta}(\qvec)]$
in real space.
For a more general source term $S_{\alpha\beta}(\rvec)$ we have of course a convolution relation
\begin{equation}
R_{\alpha\beta}(\rvec) = \frac{1}{V} 
\int \ddiff \rvec' C_{\alpha\beta\gamma\delta}(\rvec-\rvec') \ S_{\gamma\delta}(\rvec')
\label{eq_ten_ps_D}
\end{equation}
which reduces to Eq.~(\ref{eq_ten_ps_C}) for a point source.
Importantly, all statements made above for the reciprocal space remain valid in real space,
i.e. that $R_{\alpha\beta}(\rvec)$ is an isotropic tensor field if
$C_{\alpha\beta\gamma\delta}(\rvec)$ and $S_{\alpha\beta}(\rvec)$ are isotropic
and, more specifically, that $R_{12}(\rvec)$ is a quadrupolar field while
$C_{1212}(\rvec)$ is octupolar.

Importantly, in many physical situations the source is in fact {\em not} isotropic
and thus in turn the response field {\em not} consistent with Eq.~(\ref{eq_ten_field_o2}).
We remind that, e.g., the mechanical response of amorphous solids under loading proceeds from 
local and irreversible rearrangements, resetting disorder locally thus generating a highly 
non-trivial mechanical noise (``shear transformation zones") 
\cite{Argon79,Bocquet04,Tanguy11,Barrat18,Langer98}.
According to a popular model of localized plastic failure two orthogonal twin force 
dipoles of {\em opposite} signs may be imposed at the origin \cite{Bocquet04}.
This suggests to relax the isotropy condition for $s_{\alpha\beta}$.
Since the source tensor is still symmetric it may be diagonalized by an appropriate rotation 
of the coordinate system where $s_{12}=s_{21}=0$ and $s_{11}$ and $s_{22}$ become the two
(in general not identical) eigenvalues. Hence,
\begin{equation}
R_{\alpha\beta}(\qvec) = s_{11} C_{\alpha\beta11}(\qvec) + s_{22} C_{\alpha\beta22}(\qvec)
\nonumber
\end{equation}
with the isotropic correlation tensor field still being given by Eq.~(\ref{eq_ten_Tabcd_d2_C}).
Specifically, this implies
\begin{equation}
R_{12}(\qvec) = \qhat_1 \qhat_2 \left[ i_3(q) (s_{11}+s_{22})
+ i_4(q) \underline{(s_{11} \qhat_1^2 + s_{22} \qhat_2^2)} \right].
\nonumber
\end{equation}
For $s_{11}=s_{22}$ the underlined term becomes a constant
and we recover the isotropic and quadrupolar response field discussed above.
Interestingly, for eigenvalues of opposite sign, $s_{11}=-s_{22}$, we obtain
\begin{equation}
R_{12}(\qvec) = s_{11} i_4(q) \ \qhat_1 \qhat_2 (\qhat_1^2-\qhat_2^2) \propto \sin(4\theta).
\label{eq_ten_ps_F}
\end{equation}
The (non-isotropic) response field $R_{12}(\qvec)$ thus has in this case the same multipole pattern 
as the (isotropic) correlation field $C_{1212}(\qvec)$ albeit shifted by an angle $\pi/8$
\cite{foot_multipolar}.
It is readily seen by inverse FT that the same behavior applies in real space.

In summary, 
two different types of angular dependence of a response field must be distinguished. 
If the angular dependence is consistent with Eq.~(\ref{eq_ten_field_o2}),
this behavior should not be called ``anisotropy" since the angle dependence is basically
a trivial consequence of the fact that tensor field components are measured.
If on the other hand $R_{\alpha\beta}(\qvec)$ is not consistent with Eq.~(\ref{eq_ten_field_o2})
this suggests that either the correlation tensor field $C_{\alpha\beta\gamma\delta}(\qvec)$
and/or the source tensor $s_{\alpha\beta}$ are not isotropic. In many physical situations
this is in fact expected for the source term while the correlation fields may be assumed
to be isotropic. The physical behavior of response and correlation fields,
albeit closely related, then differ and should thus not be lumped together.

\section{Computational issues}
\label{algo}

\paragraph*{Algorithm and systems.}
We present below numerical data obtained for amorphous glasses formed by 
polydisperse Lennard-Jones (pLJ) particles \cite{spmP1,lyuda22a} 
simulated by means of Monte Carlo (MC) simulations \cite{AllenTildesleyBook}.
See Appendix~\ref{comp} for details (Hamiltonian, units, configuration preparation, data averaging). 
We focus on systems containing $n=10000$ particles at a working temperature $T=0.2$
much lower than the glass transition temperature $\Tglass \approx 0.26$,
i.e. for our largest sampling time $\tsamp = 10^7$ the systems behave as solid
elastic bodies and all stochastic processes are stationary \cite{spmP1}.
Importantly, all $\Nc=200$ completely independent configurations $c$ 
are quenched and tempered by means of a mix of local and swap MC hopping moves \cite{Berthier17,spmP1}
(being thus effectively kept adiabatically at thermal equilibrium) 
while the data production runs are sampled only using local MC moves.
For each $c$ we store several time-series containing each $\Nt=10000$ frames $t$. 


\begin{figure}[t]
\centerline{\resizebox{.80\columnwidth}{!}{\includegraphics*{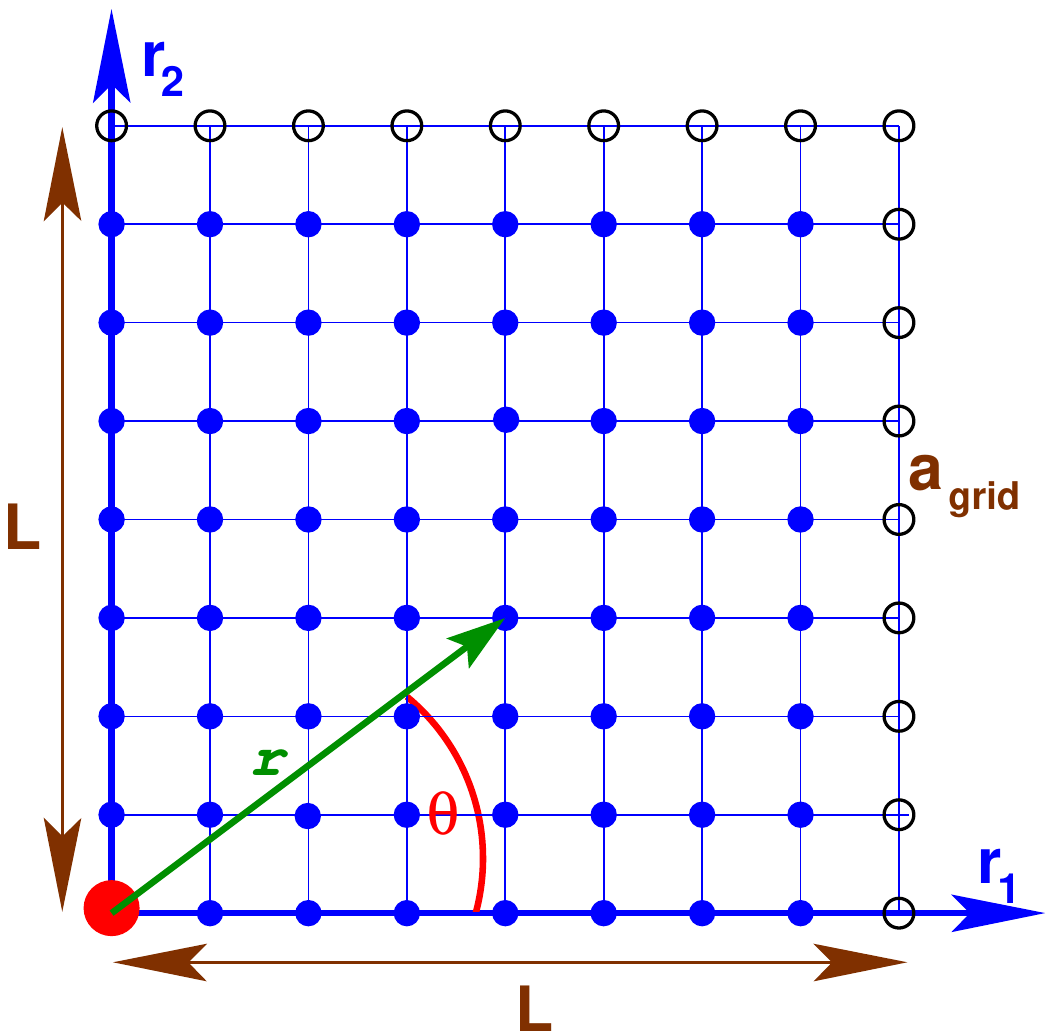}}}
\caption{Two-dimensional ($d=2$) square lattice with $\agrid$ being the lattice constant
and $\nLgrid=L/\agrid$ the number of grid points in one spatial dimension.
The filled circles indicate microcells of the principal box,
the open circles some periodic images.
The spatial position $\rvec$ of a microcell is either given by the $r_1$- and
$r_2$-coordinates (in the principal box) or by the distance $r=|\rvec|$
from the origin (large circle) and the angle $\theta$.
}
\label{fig_grid}
\end{figure}

\paragraph*{Data sampling and analysis.}
In a first step various {\em instantaneous} properties are computed for each $t$ 
which are then ``$t$-averaged" over the correlated $t$
and finally ``$c$-averaged" over the independent $c$.
We thus characterize, e.g., the elastic modulus tensor $\Eabcd$
by means of the stress-fluctuation formalism \cite{Lutsko88,Lutsko89,WXP13,WXP13c,WXB15,WXBB15}
from which a finite Young modulus $E \approx 45$ is obtained (cf. Sec.~\ref{e2E_viscoelast}).
Similarly, we compute in turn 
\begin{itemize}
\item
for each $c$ and $t$ the stress tensor field $\sigab(\rvec,t)|_c$ (cf. Appendix~\ref{stress})
using a regular square grid as shown in Fig.~\ref{fig_grid}
with a lattice constant $\agrid \approx 0.2$,
\item
the $t$-averaged fields $\Sab(\rvec)|_c$,
\item
by Fast-Fourier transform the corresponding stress fields 
$\Sab(\qvec)|_c=\Fcal[\Sab(\rvec)|_c]$ in reciprocal space, 
\item
the correlation functions $\Cabcd(\qvec)|_c$ in reciprocal space for each configuration $c$,
cf.~Eq.~(\ref{eq_comp_Cabcd_c}),
\item
the $c$-average $\Cabcd(\qvec)$, cf.~Eq.~(\ref{eq_comp_caver_Cabcd}), 
\item
and finally by inverse FT $\Cabcd(\rvec) = \Fcal^{-1}[\Cabcd(\qvec)]$
the correlation functions in real space.
\end{itemize}
To obtain the correlation functions in rotated coordinates
we rotate first $\Sab(\qvec)|_c \to \Sab^{\prime}(\qvec)|_c$ and perform then all the subsequent steps as before.
We note finally that 
real and reciprocal space correlation functions have the {\em same} dimension
due to our FT convention (cf. Appendix~\ref{FT}).

\section{Main numerical results}
\label{res}

\begin{figure}[t]
\centerline{\resizebox{0.9\columnwidth}{!}{\includegraphics*{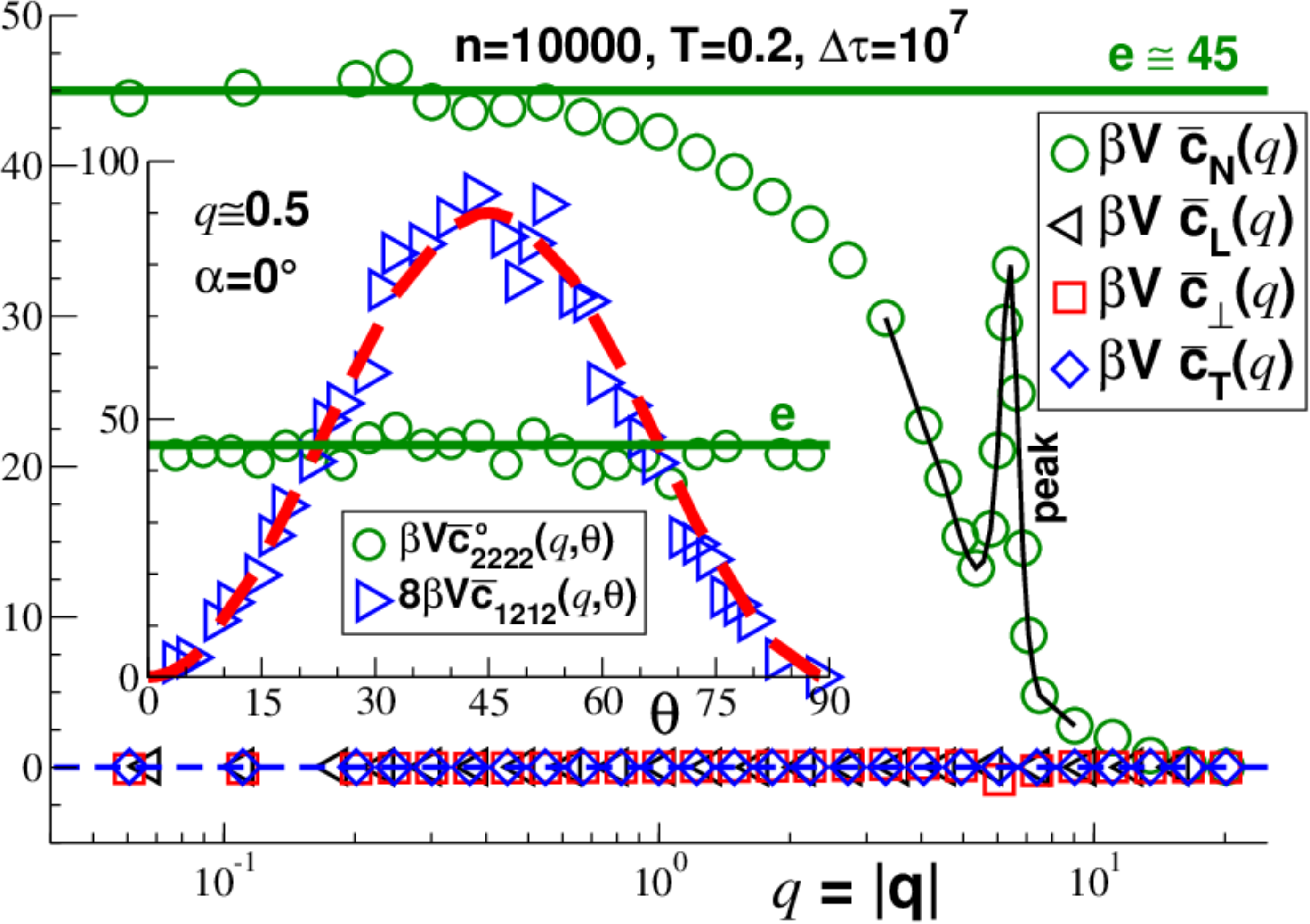}}}
\caption{ICFs $\Ctwo(q)$, $\Cone(q)$, $\Cthree(q)$ and $\Cfour(q)$ for a large sampling time $\tsamp=10^7$ 
and $n=10000$ particles.
Horizontal solid lines indicate the phenomenological constant $\Etilde\approx 45$.
Inset: 
$\bar{c}_{2222}^{\circ}(q,\theta)$ does not depend on $\theta$
while $\Cxyxy(q,\theta)$ is well described by Eq.~(\ref{eq_res_Cxyxy_q_E}) (dashed line).
Main panel: 
$\Cone(q)$, $\Cthree(q)$ and $\Cfour(q)$ vanish 
while $\beta V \Ctwo(q) \simeq \Etilde > 0$ for $q \ll 1$.
A thin solid line emphasizes the peak of $\Ctwo(q)$ at $q \approx 6.5$.
}
\label{fig_Crot_q}
\end{figure}

\paragraph*{Determination of ICFs in NRC.}
Before we shall have a closer look at $\Cabcd(\qvec)$ in standard unrotated or rotated coordinates
let us first characterize the correlations in NRC, i.e. for {\em each} wavevector $\qvec$
the coordinate system is rotated until the $1$-axis coincides with the $\qvec$-direction.
We mark these new tensor field components by ``$\circ$"
to distinguish them from standard rotated tensor field components (marked by primes ``$\prime$")
using the {\em same} rotation for all $\qvec$.
Note that $q_{\alpha}^{\circ} = q \delta_{1\alpha}$.
The ICFs $\Cabcd^{\circ}(\qvec) =
\langle \bar{\sigma}^{\circ}_{\alpha \beta}(\qvec) \bar{\sigma}^{\circ}_{\gamma \delta}(-\qvec)\rangle$
are thus obtained using the ``Invariant Stress Fields" (ISFs)
$\bar{\sigma}^{\circ}_{\alpha\beta}(\qvec)|_c$ rotated {\em differently} for each $\qvec$.
Importantly, for strictly isotropic systems $\Cabcd^{\circ}(\qvec)$
only depends on the magnitude $q$ of $\qvec$ but not on its direction $\qhatvec$.
Consistently with Eq.~(\ref{eq_ten_Tabcd_d2_A}) and following Ref.~\cite{lyuda18}, we define
\begin{eqnarray}
\Cone(q)   & \equiv & \la \bar{c}^{\circ}_{1111}(\qvec) \ra_{\qhatvec}, \label{eq_res_ICF_def} \\
\Ctwo(q)   & \equiv & \la \bar{c}^{\circ}_{2222}(\qvec)\ra_{\qhatvec}, \nonumber \\
\Cthree(q) & \equiv & \la \bar{c}^{\circ}_{1122}(\qvec) \ra_{\qhatvec} \mbox{ and } \nonumber \\
\Cfour(q)  & \equiv & \la \bar{c}^{\circ}_{1212}(\qvec) \ra_{\qhatvec} \nonumber
\end{eqnarray}
where we average over all $\qhatvec$ with $|\qvec| \approx q$
(using a bin width similar to the lattice spacing of the grid in reciprocal space).
The $\qhatvec$-averaged four ICFs are shown in the main panel of Fig.~\ref{fig_Crot_q}
for our largest sampling time $\tsamp$.
The central observation is that $\Cone(q)$, $\Cthree(q)$ and $\Cfour(q)$
vanish for sufficiently large $\tsamp$ while $\Ctwo(q)$ remains finite.
Moreover, as emphasized by the bold solid line in the main panel
\begin{equation}
\beta V \Ctwo(q) \simeq \Etilde \approx 45 \mbox{ for } q \ll 1
\label{eq_res_Ctwo2E}
\end{equation}
with $\beta=1/T$ being the inverse temperature, $V$ the system volume
and $\Etilde$ a phenomenological constant characterizing
the plateau in the hydrodynamic limit.
As shown in the inset of Fig.~\ref{fig_Crot_q} for $\bar{c}^{\circ}_{2222}(q,\theta)$
and $\alpha=0^{\circ}$, we have checked for several tensor field components 
$\bar{c}^{\circ}_{\alpha\beta\gamma\delta}(q,\theta)$ that the expected $\theta$-independence
for isotropic systems holds (within statistical accuracy).


\begin{figure}[t]
\centerline{\resizebox{0.9\columnwidth}{!}{\includegraphics*{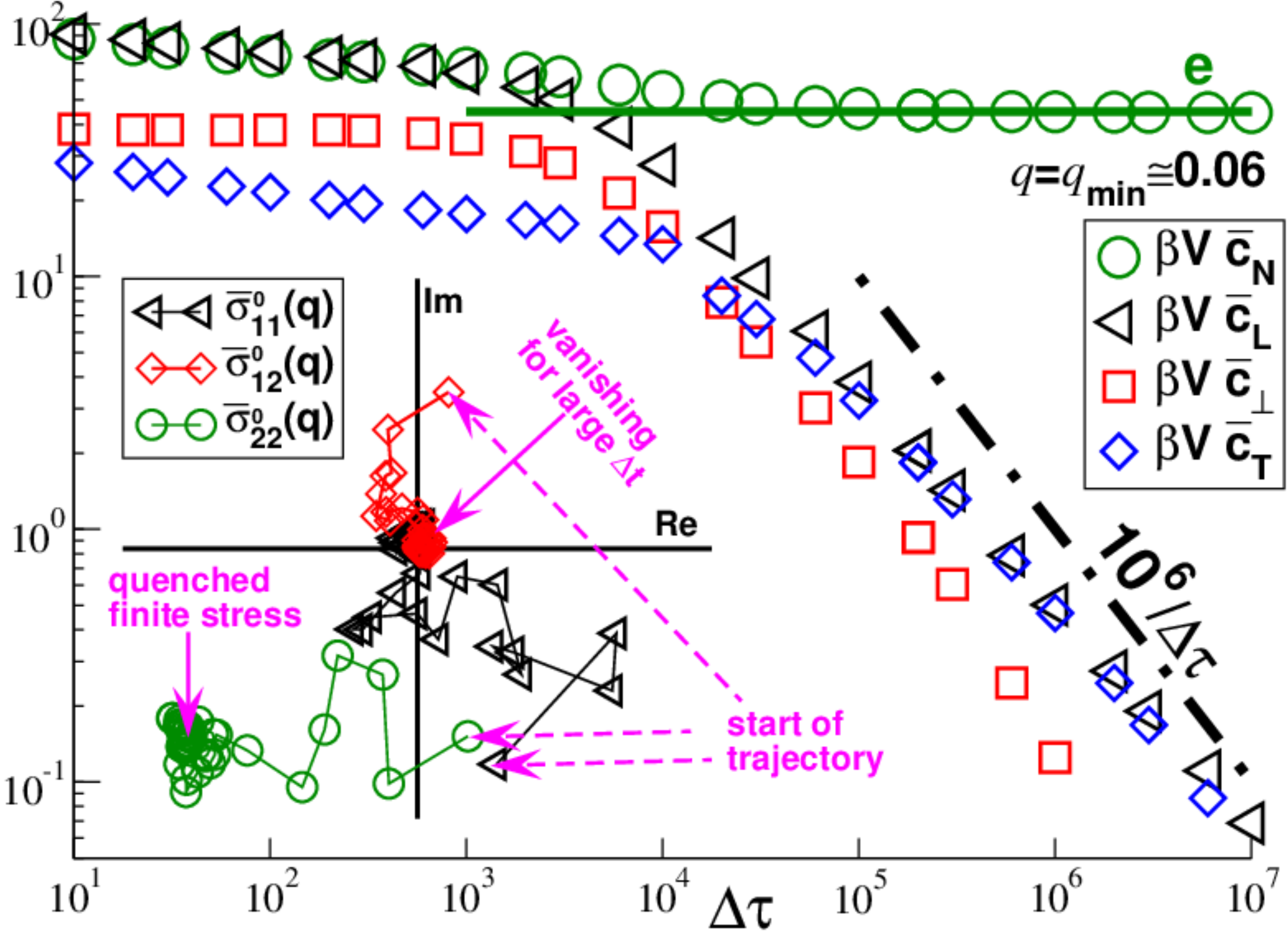}}}
\caption{$\tsamp$-dependence in reciprocal space using NRC and $q\approx 0.06$.
Main panel: $\tsamp$-dependence of ICFs demonstrating that $\Ctwo(q,\tsamp)$ remains finite 
while all other ICFs decay inversely with $\tsamp$.
Inset: $\ISFlongi(\qvec,\tsamp)$, $\ISFshear(\qvec,\tsamp)$ and $\ISFnorm(\qvec,\tsamp)$
in the complex plane (horizontal axis for real part, vertical axis for imaginary part)
as a function of $\tsamp$ (using a logarithmic scale for the data points)
for one $\qvec$ and one $c$.
$\ISFlongi(\qvec,\tsamp)$ and $\ISFshear(\qvec,\tsamp)$ vanish for large $\tsamp$, 
i.e. converge to the origin of the complex plane,
while $\ISFnorm(\qvec,\tsamp)$ has a {\em finite} attractor $\ISFnormQ(\qvec)$.
}
\label{fig_sigrot_tsamp}
\end{figure}

\paragraph*{Sampling time dependence and symmetry breaking.}
The dependence of the ICFs on $\tsamp$ is summarized in Fig.~\ref{fig_sigrot_tsamp}.
Let us first note that the ($t$-averaged) force
$\bar{g}_{\alpha}(\rvec) = \partial_{\beta} \bar{\sigma}_{\alpha\beta}(\rvec)$
acting on each material element becomes in reciprocal space 
\begin{equation}
\bar{g}^{\circ}_{\alpha}(\qvec) = i q^{\circ}_{\beta} \ \bar{\sigma}^{\circ}_{\alpha\beta}(\qvec)
\label{eq_forceNRC}
\end{equation}
using NRC.
Since $q^{\circ}_1 = q$ is finite (for $\qvec \ne \bfzero$),
finite $\bar{\sigma}^{\circ}_{11}(\qvec)$ and $\bar{\sigma}^{\circ}_{12}(\qvec)$
correspond to finite forces $\bar{g}^{\circ}_{\alpha}(\qvec)$ which in turn generate fluxes.
Finite $\ISFlongi(\qvec,\tsamp)$ and $\ISFshear(\qvec,\tsamp)$ must thus rapidly vanish,
as seen from the corresponding trajectories in the inset of Fig.~\ref{fig_sigrot_tsamp},
and therefore $\Cone(q,\tsamp)$, $\Cthree(q,\tsamp)$ and $\Cfour(q,\tsamp)$ also vanish
for large $\tsamp$ (main panel) \cite{lyuda22a}.
Since $q^{\circ}_2=0$ for all $\qvec$ the normal stress $\ISFnorm(\qvec,\tsamp)$ 
may be finite without violating static mechanical equilibrium. As shown in the inset,
the $\ISFnorm(\qvec,\tsamp)$ thus have in general {\em finite} attractors $\ISFnormQ(\qvec)$.
As emphasized by the superscript ``q", these are for realistic $\tsamp$ essentially quenched stresses. 
(Only for $\tsamp$ of order of the $\alpha$-relaxation time $\taualph$ these attractors become
weakly time-dependent diffusively decaying for symmetry reasons towards the origin of the complex plane.
Note that $\tsamp \ll \taualph$ for $T=0.2$.)
Since the large-$\tsamp$ limit of $\Ctwo(q,\tsamp)$ is the typical squared magnitude of the 
$\ISFnormQ(\qvec)$ in the complex plane, this implies 
\begin{equation}
\Etilde = \beta V \la \ISFnormQ(\qvec) \ISFnormQ(-\qvec) \ra.
\label{eq_ISFnormQ2Etilde}
\end{equation}
This means that $\Etilde$ is a finite $\tsamp$-independent static property characterizing 
the typical size of the (continuous) symmetry breaking associated with the stress components normal to the wavevectors
for each configuration $c$. Note that $\Etilde$ does thus not depend on whether we use, e.g., 
a momentum conserving or an overdamped simulation scheme \cite{AllenTildesleyBook}.
In general, it is a fitting parameter depending on the distribution of the 
$\ISFnormQ(\qvec)$ caused by the preparation history.
Interestingly, it is observed that $\Etilde$ is similar to the Young modulus $E$.
This result is in fact expected from recent studies on equilibrium viscoelastic fluids 
(including supercooled liquids and glasses) \cite{lyuda18,lyuda22a}
showing that the ICFs may be expressed in the small-$q$ limit
in terms of invariant macroscopic relaxation functions 
\cite{RubinsteinBook,lyuda18,lyuda22a,Fuchs17,Fuchs18,Fuchs19}.
Naturally, this requires additional assumptions
the crucial point being here that the systems must be at thermal equilibrium.
See Sec.~\ref{e2E} for more details.
%


\paragraph*{Reciprocal space correlation functions.}
We turn now to a coordinate system rotated 
as in Fig.~\ref{fig_intro}(b) by the {\em same} angle $\alpha$ for all $\qvec$.
As shown in Sec.~\ref{ten_Tabcd_d2} and using the form-invariance of isotropic
tensor fields, cf. Eq.~(\ref{eq_ten_iso_field}),
the four ICFs determine in $d=2$ the isotropic forth-order tensor field \cite{lyuda18}
\begin{eqnarray}
\Cabcdprime(\qvec) & = & \left[\Ctwo-2\Cfour\right] 
 \ \ \delta_{\alpha\beta} \delta_{\gamma\delta} \label{eq_res_Cabcd_general} \\
& + & \Cfour \ \left(\delta_{\alpha\gamma}\delta_{\beta\delta}+
              \delta_{\alpha\delta}\delta_{\beta\gamma} \right) \nonumber \\
& + & \left[\Cthree-\Ctwo + 2\Cfour \right] 
\ \left( \hat{q}^{\prime}_{\alpha} \hat{q}^{\prime}_{\beta} \delta_{\gamma\delta}+  
\hat{q}^{\prime}_{\gamma} \hat{q}^{\prime}_{\delta} \delta_{\alpha\beta} \right) 
			      \nonumber \\
& + & \left[\Cone+\Ctwo-2\Cthree-4\Cfour \right]
\ \ \hat{q}^{\prime}_{\alpha} \hat{q}^{\prime}_{\beta} \hat{q}^{\prime}_{\gamma} \hat{q}^{\prime}_{\delta} \nonumber
\end{eqnarray}
with $\qhat^{\prime}_{\alpha}$ being the $\alpha$-component of $\qhatvec^{\prime}=\qhatvec$
in rotated coordinates.
The $\alpha$-rotation changes the terms in the last two lines of Eq.~(\ref{eq_res_Cabcd_general}).
(Minor generalizations are needed for $d > 2$.)
Using the known values of the ICFs of our system Eq.~(\ref{eq_res_Cabcd_general}) reduces to 
\begin{eqnarray}
\beta V \Cabcdprime(\qvec)
& \simeq & \Etilde \times \label{eq_res_Cabcd_q_E} \\
(\delta_{\alpha\beta}\delta_{\gamma\delta} 
& -&  \qhat^{\prime}_{\alpha} \qhat^{\prime}_{\beta} \delta_{\gamma\delta} - 
\qhat^{\prime}_{\gamma} \qhat^{\prime}_{\delta} \delta_{\alpha\beta} 
+ \qhat^{\prime}_{\alpha} \qhat^{\prime}_{\beta} \qhat^{\prime}_{\gamma} \qhat^{\prime}_{\delta})
\nonumber
\end{eqnarray}
for $q \ll 1$ and large $\tsamp$.
According to Eq.~(\ref{eq_res_Cabcd_q_E}) we thus
obtain, e.g., for the shear-stress autocorrelation function
\begin{equation}
8 \beta V \Cxyxyprime(\qvec) = \Etilde \ 
(1-\cos[4 (\theta-\alpha)])
\label{eq_res_Cxyxy_q_E} 
\end{equation}
for all $\alpha$.
As seen from the inset of Fig.~\ref{fig_Crot_q} for $\alpha=0^{\circ}$,
this prediction (bold dashed line) agrees perfectly with our data (triangles).


\begin{figure}[t]
\centerline{\resizebox{0.9\columnwidth}{!}{\includegraphics*{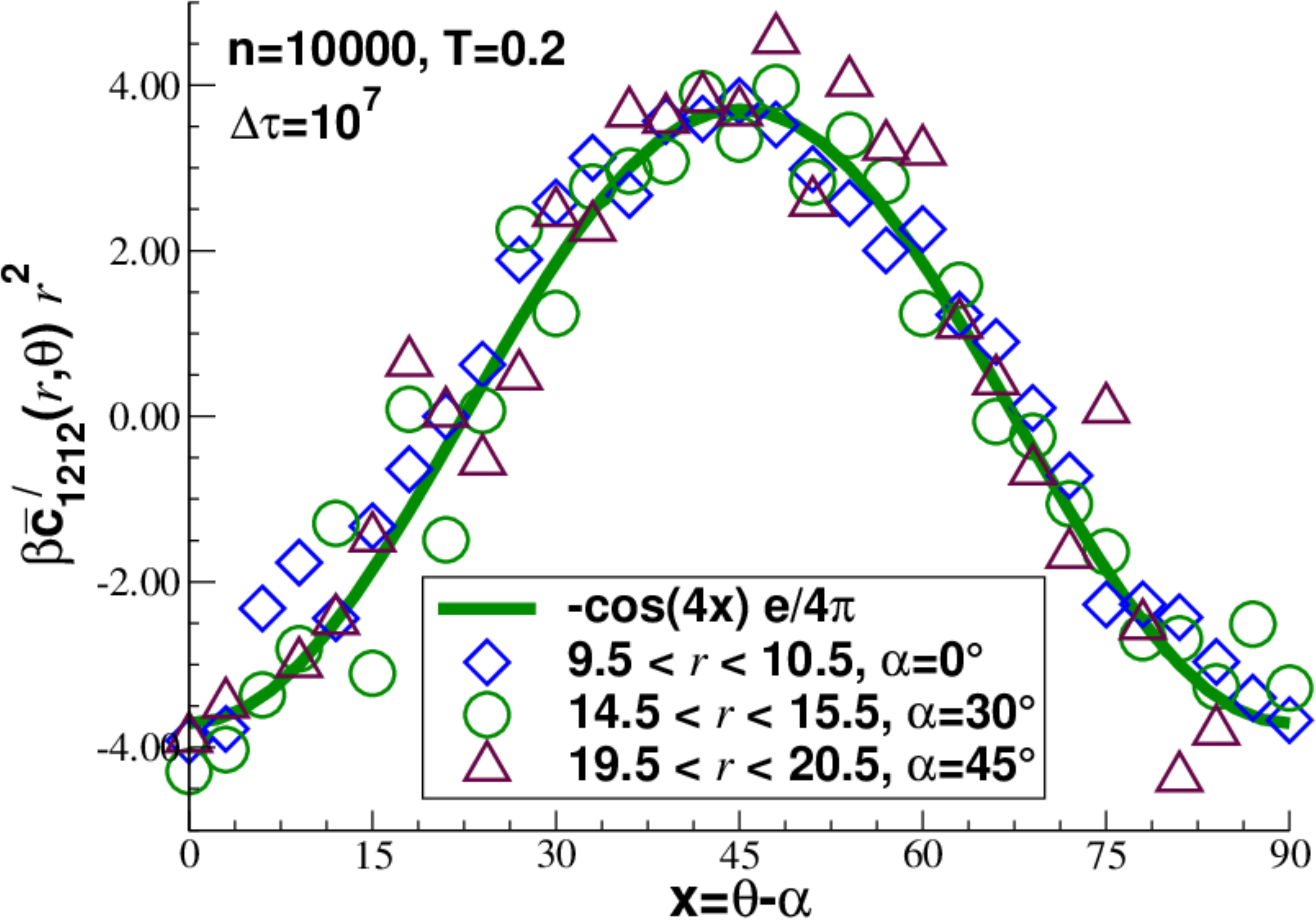}}}
\caption{Rescaled shear-stress autocorrelation function $\beta \Cxyxyprime(r,\theta) r^2$ 
as a function of $x=\theta-\alpha$ comparing data for different $r$ and $\alpha$ 
with the prediction (bold solid line).
}
\label{fig_Cabcd_thetar}
\end{figure}

\paragraph*{Real space correlation functions.}
We return now to the correlations $\Cabcdprime(\rvec) = \Fcal^{-1}[\Cabcdprime(\qvec)]$ in real space.
As already stated in the Introduction, cf. Eq.~(\ref{eq_Cxyxy_intro}),
inverse Fourier transformation (cf. Appendix~\ref{q2r}) implies
\begin{equation}
-\beta \Cxyxyprime(\rvec) \simeq \frac{\Etilde}{4\pi r^2} \cos[4(\theta-\alpha)] \mbox{ for } r \gg 1
\label{eq_res_Cxyxy_rtheta}
\end{equation}
with $\theta$ being the angle indicated in Fig.~\ref{fig_intro}(a).
The same large-$\tsamp$ limit holds for $-\beta \Cxxyyprime(\rvec)$ and for
$\beta (\Cxxxxprime(\rvec)+\Cyyyyprime(\rvec))/2$.
Moreover,
\begin{equation}
\beta (\Cxxxxprime(\rvec)-\Cyyyyprime(\rvec))/2 \simeq 2 \frac{\Etilde}{4\pi r^2} \cos[2(\theta-\alpha)]
\label{eq_res_Cdiff_rtheta}
\end{equation}
for $r \gg 1$ and for large $\tsamp$.
The angle dependence for the shear-stress autocorrelation function in real space is investigated 
in Fig.~\ref{fig_Cabcd_thetar} where we plot using linear coordinates 
$\beta \Cxyxyprime(r,\theta) r^2$ 
as a function of $x=\theta-\alpha$ for different $r$ and $\alpha$.
The data compare well with the prediction, Eq.~(\ref{eq_res_Cxyxy_rtheta}).
Naturally, the statistics deteriorates with increasing $r$.


\begin{figure}[t]
\centerline{\resizebox{0.90\columnwidth}{!}{\includegraphics*{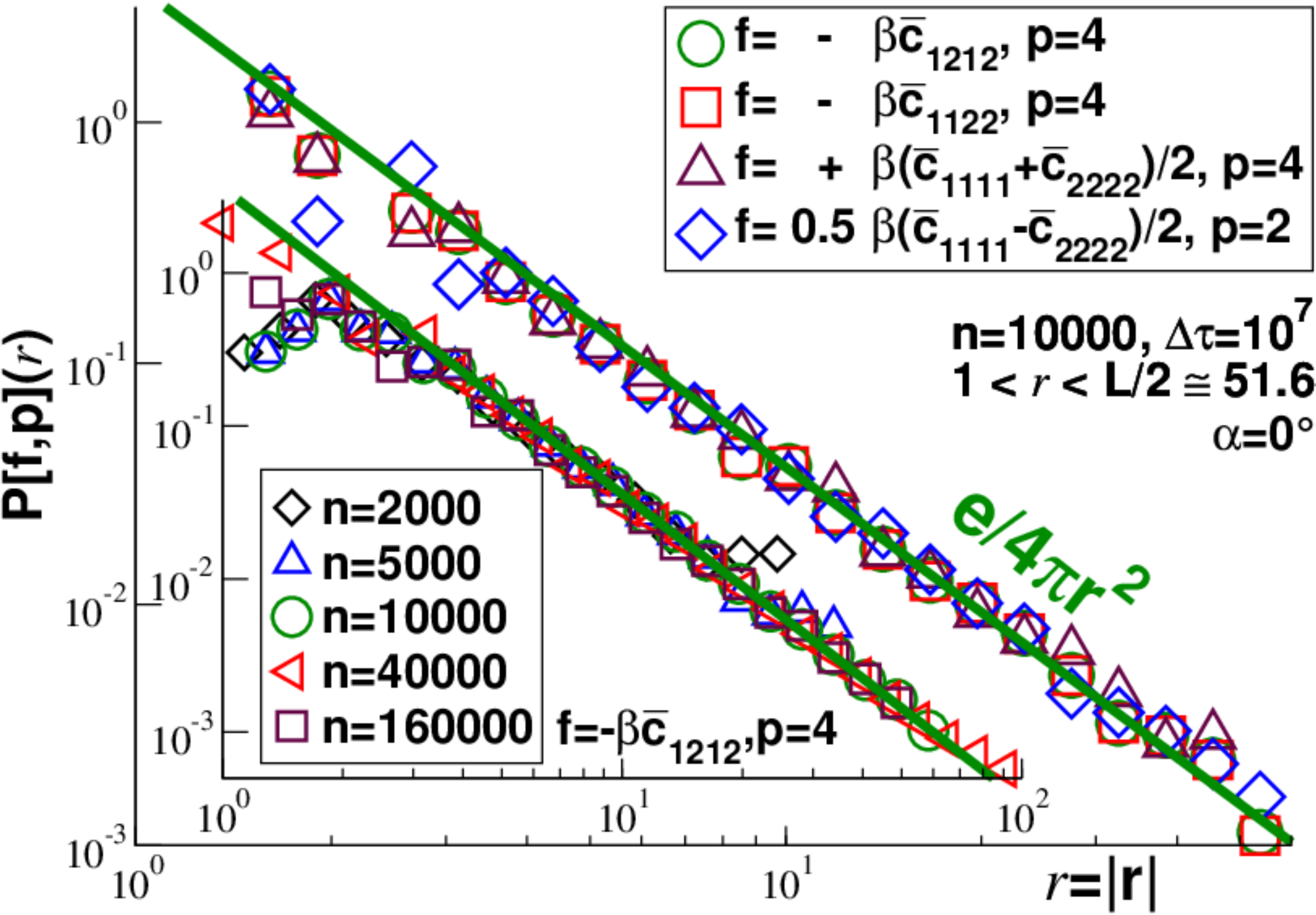}}}
\caption{$P[f(\rvec),p](r)$ for rescaled correlation functions
$f(\rvec)$ and mode number $p$ as indicated ($\alpha=0^{\circ}$).
A double-logarithmic representation is used. Only data for $1 < r < L/2$ are given.
The bold solid lines represent  $\Etilde/4\pi r^2$.
Main panel: Various projections for $n=10000$.
Inset: Projection for  $f(\rvec)=-\beta\Cxyxy(\rvec)$ and $p=4$ for different particle numbers $n$.
}
\label{fig_Cabcd_rproj}
\end{figure}

\paragraph*{Asymptotic $r$-dependence.}
A more precise check of the $r$-dependence is obtained using the $\theta$-average 
\begin{equation}
P[f,p](r) \equiv
\frac{2}{\pi} \int_0^{\pi} \ddiff \theta \ f(r,\theta) \cos(p \theta)
\end{equation}
for $p=2$ and $p=4$.
For convenience the prefactor of the integral is chosen such that
$P[\cos(2\theta),2]=P[\cos(4\theta),4]= 1$.
(On the discrete grid the integral is replaced by the sum over all grid points in
a distance interval $[r-\delta r/2,r+\delta r/2]$ which is finally normalized
by {\em half} the number of grid points.)
All correlation functions $f(\rvec)$ presented in Fig.~\ref{fig_Cabcd_rproj} are rescaled
to make their projections collapse on the same power law $\Etilde/4\pi r^2$ (bold solid lines).
The main panel presents different correlation functions for $n=10000$.
The negative signs for $\Cxyxy$ and $\Cxxyy$ are implied by Eq.~(\ref{eq_res_Cxyxy_rtheta}).
Consistently with Eq.~(\ref{eq_res_Cdiff_rtheta}) 
the projection for $(\Cxxxx-\Cyyyy)/2$ is additionally rescaled with a prefactor $\beta/2$.
Focusing on $f(\rvec)=-\beta \Cxyxy(\rvec)$ and $p=4$ we verify in the inset
the system-size independence of these results.
Data for a broad range of particle numbers $n$ are presented.
As can be seen, all data nicely collapse on $\Etilde/4\pi r^2$, confirming thus the
predicted long-range correlations for asymptotically large simulation boxes.
Similar results (not shown) have been found for the projections of other $\Cabcd(\rvec)$.

\section{Why and when $\Etilde \approx E$ holds}
\label{e2E}

\subsection{Introduction}
\label{e2E_intro}

The phenomenological parameter $\Etilde$ was defined in Sec.~\ref{res}
by the limit Eq.~(\ref{eq_res_Ctwo2E}).
Note that $\Etilde$ has the same dimension energy/volume as the stress or an elastic modulus.
We have verified (cf. Fig.~\ref{fig_sigrot_tsamp}) that $\Etilde$ indeed becomes $\tsamp$-independent 
for sufficiently large sampling times $\tsamp$ for our pLJ particle glasses.
Having fitted the value $\Etilde \approx 45$ and using that the ICFs 
$\Cone(q)$, $\Cthree(q)$ and $\Cfour(q)$ vanish,
all numerical results for sufficiently large $\tsamp$, small $q$ or large $r$
can be explained without any additional physical insight.
We have observed, however, that $\Etilde$ is similar to the (equilibrium static) Young modulus $E$.
Albeit not strictly necessary for the main thrust of this work,
a demonstration that both constants should be similar or even equal
--- under to be specified assumptions and approximations --- must be an important finding 
allowing to estimate {\em apriori} the angular dependence of the correlation functions.
This can indeed be done following Refs.~\cite{lyuda18,lyuda22a} and 
in agreement with a different and complementary approach developed in Refs.~\cite{Fuchs17,Fuchs18,Fuchs19}.
Naturally, this requires additional physical input.
We remind first in Sec.~\ref{e2E_tsamp} how the correlation functions 
$\bar{c}(\tsamp)$ of time-averaged fields, the focus of the present work, are related to the 
correlation functions $c(\tau)$ of instantaneous fields under the assumption that the relevant 
stochastic processes are {\em stationary} \cite{foot_t_tau}.
A short recap of linear viscoelasticity is given in Sec.~\ref{e2E_viscoelast}. 
Using the general theoretical predictions for the ICFs of instantaneous stress fields 
\cite{lyuda18,lyuda22a}, reminded in Sec.~\ref{e2E_ICFs},
it is shown (cf. Sec.~\ref{e2E_ICFs_tlong}) that $\Etilde \approx E$ for 
{\em equilibrated} viscoelastic fluids with a sufficiently broad elastic plateau.

\subsection{Instantaneous and time-averaged fields}
\label{e2E_tsamp}

\paragraph*{General connection for stationary processes.}
We discuss in this work spatial correlation functions $\bar{c}$ (both in real and 
reciprocal space) of time-averaged stress fields computed over $\Nt = \tsamp/\tincr$ 
instantaneous configurations. 
The various correlation functions thus depend in principle on the sampling time $\tsamp$
as we have seen in Fig.~\ref{fig_sigrot_tsamp}. 
As shown elsewhere \cite{spmP1} 
assuming a {\em stationary} stochastic process
(both for global properties as for fields) 
the $\tsamp$-dependence of $\bar{c}(\tsamp)$ can be traced back via
\begin{equation}
\bar{c}(\tsamp) = \frac{2}{\tsamp^2} \int_0^{\tsamp} \ddiff \tau \ (\tsamp-\tau) \ c(\tau)
\label{eq_e2E_tsamp_Ctsamp}
\end{equation}
to the time dependence of the corresponding correlation function $c(\tau)$ of the instantaneous fields. 
Please note that Eq.~(\ref{eq_e2E_tsamp_Ctsamp}) 
is closely related to the equivalence of the Einstein relation, 
corresponding to $\bar{c}(\tsamp)$,
and the Green-Kubo relation, corresponding to $c(\tau)$,
for transport coefficients \cite{HansenBook,AllenTildesleyBook,spmP1}.
We thus study in this work correlations within the Einstein picture.
This has the advantage that the integral Eq.~(\ref{eq_e2E_tsamp_Ctsamp}) filters 
irrelevant high frequencies, i.e. $\bar{c}(\tsamp)$ is a natural smoothing function
of the instantaneous field correlation function $c(\tau)$
on which previous work has focused on
\cite{Fuchs17,Fuchs18,Fuchs19,lyuda18,Lemaitre2004,Lemaitre14,Lemaitre15,Lemaitre17,Lemaitre18,Harrowell16,lyuda22a,MM22}.

\paragraph*{Plateau values and asymptotic behavior.}
Obviously, Eq.~(\ref{eq_e2E_tsamp_Ctsamp}) implies that $c(\tau)$ is constant {\em iff} 
$\bar{c}(\tsamp)$ is constant and both constants are equal.
Importantly, this even holds if $c(\tau)$ and $\bar{c}(\tsamp)$ are only constant
for a finite but sufficiently large time window \cite{spmP1}.
Hence, if $c(\tau)$ becomes constant in the large-time limit
$\bar{c}(\tsamp)$ as well becomes constant, i.e.
\begin{equation}
\lim_{\tau\to \infty} c(\tau) = \lim_{\tsamp \to \infty} \bar{c}(\tsamp) \equiv \cinf
\label{eq_e2E_tsamp_Ctsamp_cinf}
\end{equation}
with $\cinf$ being the common asymptote \cite{foot_tinf_convention}.
Thus, $\Cone(q,\tsamp) \approx \Cthree(q,\tsamp) \approx \Cfour(q,\tsamp) \simeq 0$
for large $\tsamp$ implies 
$\cone(q,\tau) \approx \cthree(q,\tau) \approx \cfour(q,\tau) \simeq 0$ for large $\tau$
and {\em visa versa}.
Similarily, 
\begin{equation}
\beta V \ctwo(q,\tau) \simeq \Etilde \ \ \mbox{\Large $\Leftrightarrow$} \ \
\beta V \Ctwo(q,\tsamp) \simeq \Etilde 
\label{eq_e2E_tsamp_ctwo_tlarge}
\end{equation}
for the ICFs of the transverse normal stresses in the low-$q$ limit for, respectively, 
$\tau\to \infty$ and $\tsamp \to \infty$ \cite{foot_tinf_convention}.

\paragraph*{Generalized Maxwell model.}
It follows directly from Eq.~(\ref{eq_e2E_tsamp_Ctsamp}) for $c(\tau) = c_p \exp(-\tau/\tau_p)$ 
that \cite{spmP1}
\begin{eqnarray}
\bar{c}(\tsamp) & = & c_p D(\tsamp/\tau_p) \mbox{ with } \nonumber \\
D(x) &  = & 2 [\exp(-x)-1+x]/x^2 
\label{eq_tsamp_Debye}
\end{eqnarray}
being the ``Debye function" well known in polymer science \cite{DoiEdwardsBook,RubinsteinBook}.
For systems with overdamped dynamics, such as for our MC simulations,
one expects the relaxation dynamics to be described by a linear superposition
of exponentially decaying Maxwell modes \cite{RubinsteinBook,FerryBook}.
For such a ``Generalized Maxwell model" Eq.~(\ref{eq_tsamp_Debye}) generalizes 
to the superposition \cite{spmP1}
\begin{equation}
\bar{c}(\tsamp) = \cinf + \sum_{p=1}^{\pmax} c_p D(\tsamp/\tau_p)
\label{eq_tsamp_GMM}
\end{equation}
with $c_p$ being the amplitude and $\tau_p$ (with $0 < \tau_p < \infty$)
the relaxation time of a mode $p$.
Note that $\cinf$ corresponds to the modes with virtually infinite relaxation times.
We remind that given a sufficiently high number of modes $p$ (or a distribution of modes)
it is in principle always possible to fit any reasonable $\bar{c}(\tsamp)$ using the
standard numerical techniques \cite{FerryBook,Provencher1982,lyuda22a}.
In any case, for $\tsamp \gg \taustar$ with $\taustar=\tau_1$ being the
largest (Maxwell) relaxation time Eq.~(\ref{eq_tsamp_GMM}) leads to
\begin{equation}
\bar{c}(\tsamp) \simeq \cinf + \frac{2}{\tsamp} \sum_{p=1}^{\pmax} c_p \tau_p,
\label{eq_tsamp_GMM_tlarge}
\end{equation}
i.e. $\bar{c}(\tsamp)-\cinf$ ultimately decays inversely with $\tsamp$.
This decay is emphasized by the dashed-dotted line in Fig.~\ref{fig_sigrot_tsamp}.

\subsection{Linear elasticity and viscoelasticity}
\label{e2E_viscoelast}


\begin{figure}[t]
\centerline{\resizebox{1.0\columnwidth}{!}{\includegraphics*{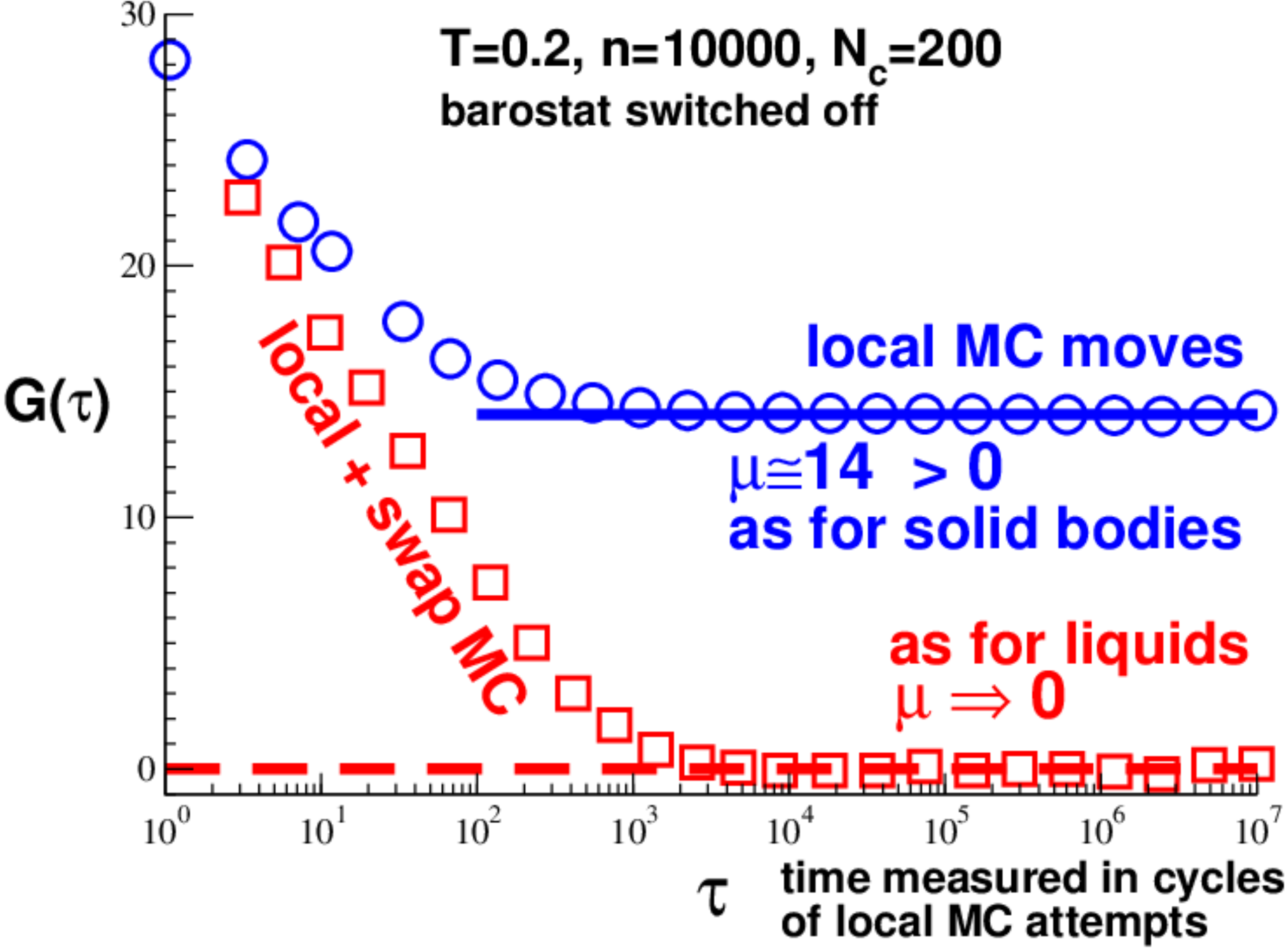}}}
\caption{Comparison of the standard MC algorithm only including local hopping moves (circles)
\cite{WXP13} with the dramatically more efficient variant including swap MC hopping moves (squares)
\cite{Berthier17,spmP1} for pLJ systems with $n=10000$ particles
focusing on the shear-stress relaxation function $G(\tau)$ at $T=0.2$.
$G(\tau)$ is computed using the shear-stress autocorrelation function and an additive constant
\cite{WXB15,WXBB15,spmP1,lyuda22a}.
$G(\tau) \approx \mu \approx 14$ for $\tau \gg 10^3$ (solid horizontal line)
if only local moves are used while $G(\tau) \to 0$ (dashed horizontal line)
if the unphysical swap MC moves are added.
Including swap MC moves for the quenching and tempering of the systems thus allows
to bring the systems close to (liquid) equilibrium at $T=0.2$. The final production runs are sampled
only using local moves.
}
\label{fig_Gt}
\end{figure}


\paragraph*{Static elastic moduli.}
Consistently with Sec.~\ref{ten_ten},
the elastic modulus tensor $\Eabcd$ for isotropic systems 
is completely described by {\em two} invariants, say the two Lam\'e moduli $\lambda$ and $\mu$.
Using Eq.~(\ref{eq_ten_ten_o4}) $\Eabcd$ may thus be written as 
\cite{LandauElasticity,TadmorCMTBook}
\begin{equation}
\Eabcd = \lambda \delta_{\alpha\beta} \delta_{\gamma\delta}
+ \mu \left(\delta_{\alpha\gamma}\delta_{\beta\delta}+
\delta_{\alpha\delta}\delta_{\beta\gamma}\right). 
\label{eq_e2E_Eabcd_iso}
\end{equation}
We have determined $\lambda$ and $\mu$ by means of the stress-fluctuation formalism
described elsewhere \cite{Lutsko88,Lutsko89,WXP13,WXP13c,WXB15,WXBB15}.
This shows that for our pLJ particle systems at $T=0.2$ we have
$\lambda \approx 39$ and $\mu \approx 14$.
The latter value is indicated by the bold horizontal line in Fig.~\ref{fig_Gt}.
(We have verified that the fluctuations of $\lambda$ and $\mu$ between different 
independent configurations $c$ are negligible.)
Alternatively, one may describe the elastic response of a body by means of
the creep compliance tensor $\Jabcd$ being the inverse of the elastic modulus tensor:
\begin{equation}
J_{\alpha\beta\gamma\delta} E_{\gamma\delta\alpha'\beta'} =
\frac{1}{2} \left(
\delta_{\alpha\alpha'}\delta_{\beta\beta'}+ \delta_{\alpha\beta'}\delta_{\alpha'\beta}
\right).
\label{eq_e2E_Jabcd_Eabcd_inv}
\end{equation}
For isotropic bodies \cite{LandauElasticity,TadmorCMTBook}
\begin{equation}
\Jabcd = 
\frac{1+\nu}{2E} \left(\delta_{\alpha\gamma} \delta_{\beta\delta}
+ \delta_{\alpha\delta} \delta_{\beta\gamma} \right)
-\frac{\nu}{E} \delta_{\alpha\beta} \delta_{\gamma\delta}
\label{eq_e2E_Jabcd_iso}
\end{equation}
with $E$ being the Young modulus and $\nu$ Poisson's ratio.
Consistently with Eq.~(\ref{eq_e2E_Jabcd_Eabcd_inv})
the two sets of invariants $(\lambda,\mu)$ and $(E,\nu)$ are related in $d$ dimensions by
\begin{eqnarray}
\nu & = & \frac{\lambda}{2\mu+\lambda (d-1)} \label{eq_e2E_Poisson_d} \\
E   & = & \lambda+2\mu -(d-1) \lambda \nu. \label{eq_e2E_Young_d} 
\end{eqnarray}
For $d=3$ the usual formulae given in the standard textbooks \cite{LandauElasticity,TadmorCMTBook} 
are recovered while 
\begin{equation}
\nu = \frac{\lambda}{\lambda+2\mu} \mbox{ and }
E= 4\mu \frac{\lambda+\mu}{\lambda+2\mu} 
\mbox{ for } d=2.
\label{eq_e2E_Young_d2}
\end{equation}
Using the known values for $\lambda$ and $\mu$ for our pLJ particle systems at $T=0.2$
this implies $E \approx 45$ and $\nu \approx 0.6$.
Please note that the values $\lambda_c$ and $\mu_c$ obtained for each independent
configuration $c$ only differ very weakly for $n > 2000$ from the $c$-ensemble averages 
$\lambda = \sum_c \lambda_c/\Nc$ and $\mu=\sum_c \mu_c/\Nc$.
We may thus determine $E$ either by first obtaining for each $c$ 
a Young modulus $E_c$ using $\lambda_c$ and $\mu_c$ and from this $E = \sum_c E_c/\Nc$ or 
by directly applying Eq.~(\ref{eq_e2E_Young_d2}) for the $c$-averages $\lambda$ and $\mu$.


\paragraph*{Stress relaxation tensor for isotropic systems.}
The static elastic modulus tensor $\Eabcd$ can be generalized in the time domain 
as discussed systematically and in more detail in the literature 
\cite{FerryBook,DoiEdwardsBook,RubinsteinBook,BalucaniZoppiBook,EvansMorrissBook}.
This is important in general for viscoelastic fluids but also for the dynamical response 
of (crystalline or amorphous) solid bodies.
The forth-order stress relaxation tensor $\Eabcd(\tau)$ characterizes the time-dependence 
of the stress increment $\delta \sigab(\tau)$ as a function of an imposed strain $\epsab(\tau)$ 
\cite{foot_t_tau}.
Generalizing Eq.~(\ref{eq_e2E_Eabcd_iso}) for isotropic systems we may write 
\begin{eqnarray}
\Eabcd(\tau) & = & M(\tau) \ \delta_{\alpha\beta} \delta_{\gamma\delta} \nonumber \\
& + & G(\tau) \ \left(\delta_{\alpha\gamma}\delta_{\beta\delta}+
\delta_{\alpha\delta}\delta_{\beta\gamma}\right)
\label{eq_e2E_Eabcdt_iso}
\end{eqnarray}
with $M(\tau)$ being the ``mixed relaxation modulus" and $G(\tau)$ the ``shear-stress relaxation modulus".
The macroscopic linear response tensor is thus fully determined by (again)
{\em two} invariant response functions.
It is useful to additionally introduce the ``longitudinal relaxation modulus" $L(\tau) = M(\tau) + 2G(\tau)$.
Note that $G(\tau) = E_{1212}(\tau)$, $M(\tau)=E_{1122}(\tau)$ and $L(\tau)=E_{1111}(\tau)=E_{2222}(\tau)$.
The relaxation modulus $\Eabcd(\tau)$ reduces (continuously) to the static modulus $\Eabcd$ 
for large times, i.e. the invariant relaxation functions become
\begin{equation}
L(\tau) \to \lambda + 2\mu, G(\tau) \to \mu, M(\tau) \to \lambda \mbox{ for } \tau \to \infty
\label{eq_e2E_LGM_larget}
\end{equation}
with $\lambda$ and $\mu$ being the known Lam\'e moduli.
That this is the case can be seen for $G(\tau)$ in Fig.~\ref{fig_Gt} for two different MC variants.
As expected for equilibrium liquids the shear modulus vanishes, $\mu=0$, for the variant 
with swap MC hopping moves (squares) used for the equilibration of our systems.
The total ensemble of independent configurations $c$ thus corresponds to a liquid system
while each configuration $c$ being confined in a metabasin if only local moves are included
is an (isotropic) elastic body.
We note for later convenience that, hence,
\begin{equation}
L(\tau) - \frac{M(\tau)^2}{L(\tau)} \to \lambda + 2 \mu - \frac{\lambda}{\lambda+2\mu} = 
4 \mu \frac{\lambda+\mu}{\lambda+2\mu},
\label{eq_e2E_LGM_larget_B}
\end{equation}
being due to Eq.~(\ref{eq_e2E_Young_d2}) equal to the Young modulus $E$ in two dimension.

\paragraph*{Experimental relevant time scales.}
Strictly speaking, $\mu=E=0$ holds for all systems including 
amorphous glasses or even standard solids, if the mathematical limit ``$\tau \to \infty$" 
is read ``for times much larger than any instrinsic relaxation time of the system" \cite{RubinsteinBook}.
In practice,
the largest ``terminal" relaxation time $\taustar$
(for glasses called ``$\alpha$-relaxation time" \cite{FerryBook}) 
for many viscoelastic systems is commonly
much larger than any reasonable typical experimental or computational measurement time $\tauexp$.
Moreover, for many viscoelastic systems the relaxation moduli are approximatively constant
over many orders of magnitude in a time window $\tau_1 \ll \tau \ll \tau_2$.
This means especially that the system can support a finite shear stress in this time window.
It is common that $\tau_1$ is given by the time $\tauA$ needed to relax the ``affine strains" 
applied at $\tau=0$ or by the relaxation time $\taubasin$ within a meta-basin
in glassy materials \cite{WXP13,WXP13c,WXB15,WXBB15,spmP1,lyuda22a} 
and $\tau_2$ by the already mentioned terminal relaxation time $\taustar$.
(Obviously, much more complicated scenarios exist 
\cite{FerryBook,RubinsteinBook,BalucaniZoppiBook}.)
We assume in the following for the simplicity of the discussion
that this {\em intermediate} plateau regime becomes very broad such that
$\tau_2 \approx \taustar$ exceeds $\tauexp$ by many orders of magnitude.
This is indeed the case, e.g., 
for the shear-stress relaxation modulus $G(\tau)$ shown in Fig.~\ref{fig_Gt} for local MC moves (circles).
The meaning of ``$\tau \to \infty$" is thus (here as elsewhere in this work)
``for times much larger than $\tau_1$ but yet much smaller than $\tau_2$".
In this sense Eq.~(\ref{eq_e2E_LGM_larget}) and Eq.~(\ref{eq_e2E_LGM_larget_B})
are still valid with $\lambda$, $\mu$ or $E$ referring now to the (finite) {\em plateau values}.
It is these plateau values which are computed by means of out-of-equilibrium
methods or by means of stress or strain fluctuation formulae 
\cite{AllenTildesleyBook}
using times series with sampling times $\tau_1 \ll \tsamp \ll \tau_2$.

\subsection{Key relations for overdamped motion}
\label{e2E_ICFs}

Previous studies \cite{lyuda18,lyuda22a,Fuchs17,Fuchs18,Fuchs19} 
have focused on the correlation functions $c^{\circ}_{\alpha\beta\gamma\delta}(\qvec,\tau)$ 
of the {\em instantaneous} ``Invariant Stress Fields" (ISFs)
$\hat{\sigma}^{\circ}_{\alpha\beta}(\qvec,\tau)$ in NRC.
Importantly, it has been shown \cite{lyuda18} that the four 
``Invariant Correlation Functions" (ICFs)
\begin{eqnarray}
\cone(q,\tau)   & \equiv & \la c_{1111}^{\circ}(\qvec,\tau) \ra_{\qhatvec}, \nonumber \\
\cthree(q,\tau) & \equiv & \la c_{1122}^{\circ}(\qvec,\tau) \ra_{\qhatvec}, \nonumber \\
\cfour(q,\tau)  & \equiv & \la c_{1212}^{\circ}(\qvec,\tau) \ra_{\qhatvec} \mbox{ and } \nonumber \\
\ctwo(q,\tau)   & \equiv & \la c_{2222}^{\circ}(\qvec,\tau) \ra_{\qhatvec}, \label{eq_e2E_ICF_def}
\end{eqnarray}
averaged over all $\qvec$ of similar magnitude $q$ (within a given bin width $\delta q$),
can be expressed for small wavevectors in terms of {\em two} independent invariant relaxation functions 
(``material functions") \cite{lyuda18} characterizing the system.
Naturally, these relations are formulated in Fourier-Laplace space.
As in Ref.~\cite{lyuda22a} we use here the modified Laplace transform 
\begin{equation}
f(s) = \Lcal[f(\tau)] = s \int_0^{\infty} \ddiff \tau \ f(\tau) e^{-\tau s},
\label{eq_e2E_ICF_LT}
\end{equation}
called ``$s$-transform" \cite{lyuda22a} or ``Laplace-Carson transform" 
\cite{RubinsteinRubinstein98},
for which due to the prefactor $s$ the original function $f(\tau)$
and its Laplace transform $f(s)$ have the same dimension. We note by
$L(s)=\Lcal[L(\tau)] = M(s) + 2 G(s)$, $M(s)=\Lcal[M(\tau)]$ and $G(s)=\Lcal[G(\tau)]$
the Laplace transforms of the invariant relaxation moduli introduced 
in Sec.~\ref{e2E_viscoelast}. 
In the present work MC simulations have been used, i.e. not a momentum conserving
dynamics as assumed in Refs.~\cite{lyuda18,lyuda22a} but an overdamped simulation
dynamics where the drift velocity of particles is proportional to the imposed body force.
Due to this the slightly modified relations become 
\begin{eqnarray}
\beta V \cone(q,s)   & = &
\frac{s L(s)}{s + q^2 L(s)/\zeta} \label{eq_e2E_ICF_cone_A} \\
\beta V \cthree(q,s) & = & \frac{s M(s)}{s + q^2 L(s)/\zeta} \label{eq_e2E_ICF_cthree_A} \\
\beta V \cfour(q,s) & = &
\frac{s G(s)}{s + q^2 G(s)/\zeta} \label{eq_e2E_ICF_cfour_A} \\
\beta V \ctwo(q,s) & = &
        L(s) - \frac{q^2 M(s)^2/\zeta}{s + q^2L(s)/\zeta} \label{eq_e2E_ICF_ctwo_A}
\end{eqnarray}
with $\zeta$ being the friction constant of the overdamped dynamics \cite{foot_headconductivity}.
All relations hold for momentum conserving dynamics if $\zeta$ is substituted by 
the term $\rho s$ due to inertia with $\rho$ being the mass density.
For momentum conserving dynamics,
Eq.~(\ref{eq_e2E_ICF_cfour_A}) is well-established \cite{BalucaniZoppiBook,EvansMorrissBook};
its derivation based on the ``Fluctuation-Dissipation Theorem" (FDT)
\cite{DoiEdwardsBook,ForsterBook,EvansMorrissBook,ChaikinBook}
is given in Ref.~\cite{ANS12} in the context of viscoelastic hydrodynamic interactions in polymer liquids.
The first relation Eq.~(\ref{eq_e2E_ICF_cone_A}) was already mentioned in Ref.~\cite{ANS17}.
We note finally that these findings agree (where a comparison is possible)
with an independent and complementary approach to long-wavelength stress-correlations in viscoelastic 
fluids at equilibrium based on the Zwanzig-Mori projection operator formalism \cite{Fuchs17,Fuchs18,Fuchs19}.

\subsection{Long-time limit for viscoelastic plateau}
\label{e2E_ICFs_tlong}

\paragraph*{Asymptotic limit.}
As noted in Sec.~\ref{e2E_viscoelast} we focus on the long-time limit
of the four key relations where in a large time window $\tau_1 \ll \tau \ll \tau_2$
all relaxation moduli are approximatively constant. 
Due to the Laplace-Carson transform,
Eq.~(\ref{eq_e2E_ICF_LT}), Eq.~(\ref{eq_e2E_LGM_larget}) implies
\begin{equation}
L(s) \to \lambda + 2\mu, G(s) \to \mu, M(s) \to \lambda 
\label{eq_e2E_LGM_smalls}
\end{equation}
for small $s$ (with $\tau_1 \ll 1/|s| \ll \tau_2$). 
Using the final-value theorem of Laplace transforms we get 
\begin{eqnarray}
\cone(q,\tau) \approx \cthree(q,\tau) \approx \cfour(q,\tau) & \simeq & 0 \label{eq_e2E_tlarge_cone} \\
\beta V \ctwo(q,\tau) \to L(\tau)-\frac{M(\tau)^2}{L(\tau)}  & \simeq & E \label{eq_e2E_tlarge_ctwo}
\end{eqnarray}
for small $q$ and large $\tau$ \cite{foot_tinf_convention}.
(Eq.~(\ref{eq_e2E_LGM_larget_B}) was used in the last step of Eq.~(\ref{eq_e2E_tlarge_ctwo}).)
Using in addition Eq.~(\ref{eq_e2E_tsamp_Ctsamp_cinf}) or Eq.~(\ref{eq_e2E_tsamp_ctwo_tlarge}) 
we have thus demonstrated that for general viscoelastic fluids at thermal equilibrium 
the phenomenological parameter $\Etilde$, cf. Eq.~(\ref{eq_res_Ctwo2E}), 
is indeed given by the Young modulus $E$ in two dimensions. 
As mentioned above (cf.~Sec.~\ref{res}), these results are qualitatively understood
by the facts that finite $\Cone(q)$, $\Cthree(q)$ and $\Cfour(q)$ for large $\tsamp$ would 
violate mechanical equilibrium 
while a finite $\Ctwo(q)$ is possible since a finite transverse normal ISF $\ISFnorm(\qvec)$
cannot induce a force on a volume element 
---due to Eq.~(\ref{eq_forceNRC}) and $q_2^{\circ}=0$ in NRC for all $\qvec$ ---
and, hence, no deterministic flux.

\begin{figure}[t]
\centerline{\resizebox{1.0\columnwidth}{!}{\includegraphics*{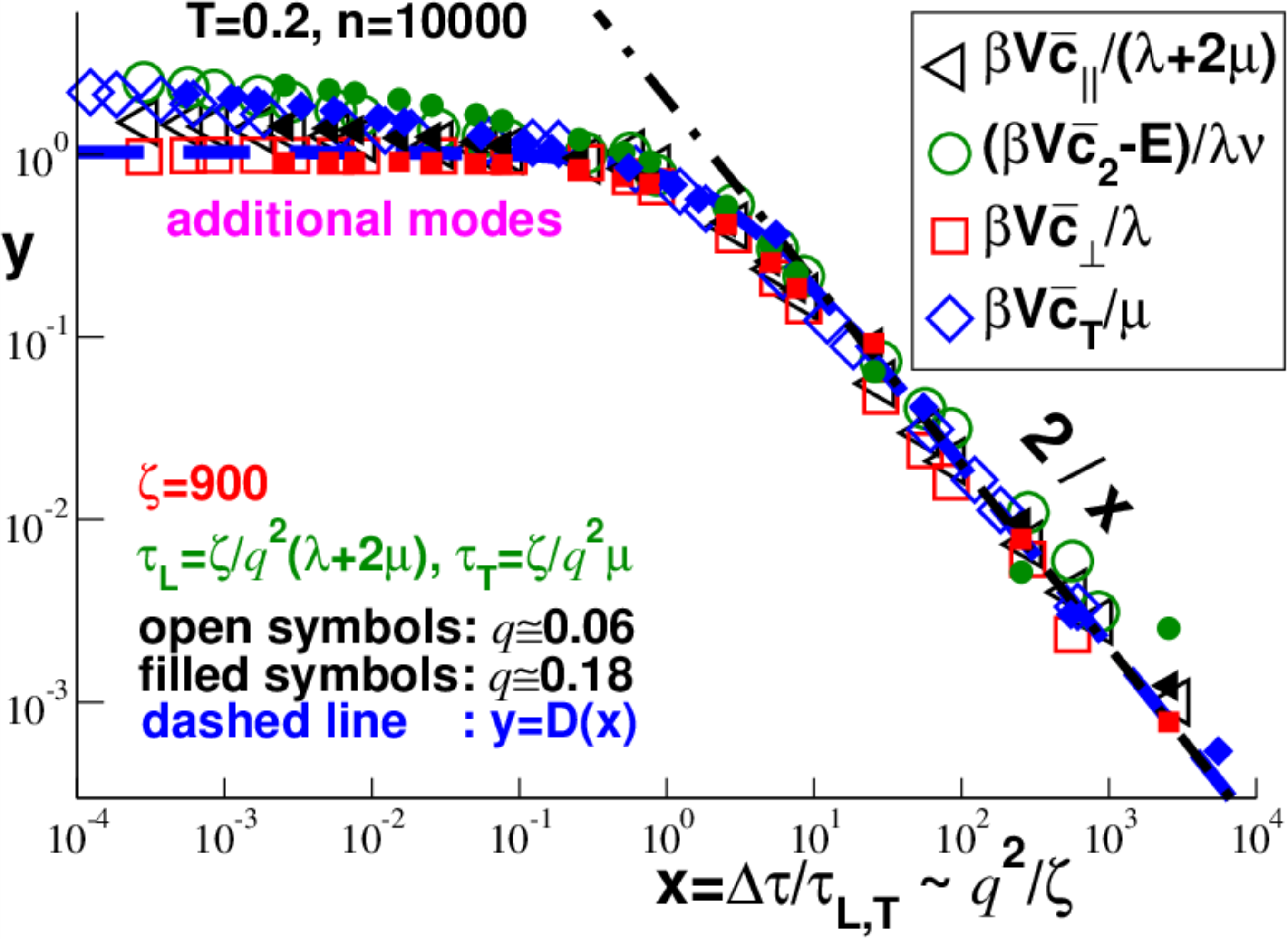}}}
\caption{Scaling of ICFs $\Cone(q,\tsamp)$,
$\Ctwo(q,\tsamp)$, $\Cthree(q,\tsamp)$ and $\Cfour(q,\tsamp)$ for large $\tsamp$.
We present the rescaled ICFs $y=\beta V\Cone/(\lambda+2\mu)$, $(\beta V\Ctwo-E)/\lambda \nu$
and $\beta V \Cthree/\lambda$ {\em vs.} $x=\tsamp/\tauL$ and
$y=\beta V \Cfour/\mu$ {\em vs.} $x=\tsamp/\tauT$
assuming $\zeta=900$ for $q=\qmin$ and $q \approx 0.18$ (filled symbols).
The bold dashed line indicates the Debye relation, the dashed-dotted line shows
the expected asymptotic power-law decay.
}
\label{fig_ICFs_tlong}
\end{figure}

\paragraph*{Predictions for leading deviations.}
Importantly, the limits Eq.~(\ref{eq_e2E_tlarge_cone}) and Eq.~(\ref{eq_e2E_tlarge_ctwo}) 
do not depend (to leading order) on the simulation dynamics,
e.g. of whether we have used a momentum conserving or overdamped dynamics. 
As emphasized by the dash-dotted line in Fig.~\ref{fig_sigrot_tsamp} 
$\Cone(\tsamp)$, $\Cthree(\tsamp)$ and $\Cfour(\tsamp)$
vanish inversely with $\tsamp$ as one expects for uncorrelated fluctuations.
We show in the remainder of this subsection that these leading deviations for finite $\tsamp$ 
are quantitatively described using Eqs.~(\ref{eq_e2E_ICF_cone_A}-\ref{eq_e2E_ICF_ctwo_A}).
To do this let us introduce the characteristic times
\begin{equation}
\tauL \equiv \zeta/q^2(\lambda+2\mu) \mbox{ and } \tauT \equiv \zeta/q^2\mu
\label{eq_ICF_tauLT}
\end{equation}
characterizing, respectively, the longitudinal and the transverse overdamped relaxation.
Using Eq.~(\ref{eq_e2E_LGM_smalls}) this leads to
\begin{eqnarray}
\beta V \cone(q,s) & \simeq & (\lambda+2\mu) \ \frac{s}{s+1/\tauL} \label{eq_ICF_cone_C} \\
\beta V \cthree(q,s) & \simeq & \lambda \ \frac{s}{s+1/\tauL} \label{eq_ICF_cthree_C} \\
\beta V \cfour(q,s) & \simeq & \mu \ \frac{s}{s+1/\tauT} \label{eq_ICF_cfour_C} \\
\beta V \ctwo(q,s) -E & \simeq &
\lambda \nu \ \frac{s}{s + 1/\tauL}.
\label{eq_ICF_ctwo_C}
\end{eqnarray}
The Laplace-Carson transformation $\Lcal[\exp(-t/\tau)] = s/(s+1/\tau)$ thus gives 
\begin{eqnarray}
\beta V \cone(q,t) & \simeq & (\lambda+2\mu) \exp(-t/\tauL) \label{eq_ICF_cone_D} \\
\beta V \cthree(q,t) & \simeq & \lambda \exp(-t/\tauL) \label{eq_ICF_cthree_D} \\
\beta V \cfour(q,t) & \simeq & \mu \exp(-t/\tauT) \label{eq_ICF_cfour_D} \\
\beta V \ctwo(q,t) -E & \simeq & \lambda \nu \exp(-t/\tauL).
\label{eq_ICF_ctwo_D}
\end{eqnarray}
for the ICFs of the instantaneous ISFs.
Hence, Eq.~(\ref{eq_tsamp_Debye}) leads to
\begin{eqnarray}
\beta V \Cone(q,\tsamp)   & \simeq & (\lambda+2\mu) \ D(\tsamp/\tauL) \label{eq_tsamp_Cone_Debye} \\
\beta V \Cthree(q,\tsamp) & \simeq & \lambda        \ D(\tsamp/\tauL) \label{eq_tsamp_Cthree_Debye} \\
\beta V \Cfour(q,\tsamp)  & \simeq & \mu            \ D(\tsamp/\tauT) \label{eq_tsamp_Cfour_Debye} \\
\beta V \Ctwo(q,\tsamp) - E & \simeq & \lambda \nu    D(\tsamp/\tauL) \label{eq_tsamp_Ctwo_Debye}
\end{eqnarray}
for the ICFs of the time-averaged ISFs.

\paragraph*{Numerical test of predicted deviations.}
These relations should allow a reasonable fit for large $\tsamp$
while additional modes may be relevant in general.
We test in Fig.~\ref{fig_ICFs_tlong} the suggested scaling for 
$q=\qmin \approx 0.06$ and $q \approx 0.18$.
The scaling variable $x$ of the horizontal axis is $\tsamp/\tauL$
for $\Cone$, $\Ctwo$ and $\Cthree$ and $\tsamp/\tauT$ for $\Cfour$.
For the scaling of $\Ctwo$ we have to subtract $E$ and, unfortunately,
the scaling function $y=(\beta V \Ctwo-E)/\lambda\nu$ (circles) is less accurate than
the scaling functions of the other three ICFs.
Using the known elastic moduli of our systems and fitting for only {\em one} parameter,
the friction coefficient $\zeta \approx 900$, we obtain for sufficiently large $x$
and small $q$ a good data collapse on the Debye function (bold dashed line).
Due to the limited system size we have only achieved good data collapse for
$\qmin \le q < 0.2$. Larger system sizes are clearly warranted in future work to
verify over more than a decade the expected $q$-dependence of the relaxation times
$\tau_{L,T}\propto 1/q^2$.
See Ref.~\cite{lyuda22a} for the discussion of other possible caveats leading to small deviations.

\section{Conclusion}
\label{conc}


\paragraph*{Summary.}
After presenting in Sec.~\ref{ten} a survey of general useful mathematical relations 
for isotropic tensor fields we have focused in Sec.~\ref{res} on the numerical description of spatial 
correlations of stress tensor fields in isotropic pLJ particle systems 
as described in Sec.~\ref{algo} and Appendix~\ref{comp}.
The phenomenological parameter $\Etilde$
was related to the Young modulus $E$ in Sec.~\ref{e2E}.
Several general results deserve to be emphasized.

\paragraph*{Angular dependences.}
As demonstrated by our computational example, 
correlation functions of tensor field components of perfectly
isotropic systems must generally depend on the components of the field 
vector and not only on its magnitude, cf. Eq.~(\ref{eq_res_Cabcd_general}).
They thus depend both on the direction of this vector in the physical system (angle $\theta$) 
and on the orientation of the coordinate system (angle $\alpha$) as shown in Fig.~\ref{fig_intro}.
Importantly, the angular dependence of all correlation functions
boils down to a dependence on the difference $\theta-\alpha$ for all $\theta$ and $\alpha$ 
as shown in panel (b) of Fig.~\ref{fig_intro} and Fig.~\ref{fig_Cabcd_thetar}. 
Obviously, 
this simple scaling (without characteristic angles) cannot hold for true anisotropic 
systems which have material functions depending explicitly on the direction of the field vector.
Unfortunately, many recent studies do not clarify, e.g., simply by rotating the coordinate system
or equivalently the orientations of the experimental devices measuring the local tensor field components, 
which of these two mathematically and physically very different types of ``anisotropy" is involved.

\paragraph*{Invariant correlation functions.}
Just as the invariance of a tensor under orthogonal transformations is only revealed 
by the full set of tensor components, one component of the correlation tensor field
may appear to be inconsistent with the assumed isotropy which becomes only 
manifested for the total set of correlation functions. 
Fortunately, this set is completely characterized by a few ($4$ in $d=2$ and $5$ in $d>2$) ICFs,
cf. Eq.~(\ref{eq_ten_field_o4_old}).
Only if a measured subset of correlation functions cannot be consistently expressed by ICFs 
a legitimate claim of ``anisotropy" can be made.
As we have seen, these ICFs are best described both theoretically 
and numerically (cf.~Fig.~\ref{fig_Crot_q}) in reciprocal space using NRC.
The investigated system was deliberately simple since
only {\em one} ICF, namely the normal stress ICF $\Ctwo(q)$, 
remains finite for large $\tsamp$ and, moreover, constant for sufficiently small $q$, 
Eq.~(\ref{eq_res_Ctwo2E}). 
This plateau in reciprocal space explains directly (cf. Appendix~\ref{q2r}) 
the long-ranged $1/r^2$-decay of the stress correlations in real space 
demonstrated in Fig.~\ref{fig_Cabcd_rproj} \cite{foot_ANS_comment}.
Only one phenomenological parameter $\Etilde$ is needed for the fitting of our data
in the large-$\tsamp$ limit, cf.~Eq.~(\ref{eq_res_Ctwo2E}).

\paragraph*{Time-averaged stress fields.}
We have studied in this work correlation functions $\Cabcd(\qvec,\tsamp)$ of time-averaged stress fields 
$\Sab(\qvec,\tsamp)$ and not the correlation functions $\cabcd(\qvec,t)$ of instantaneous fields $\sigab(\qvec,t)$. 
As reminded in Sec.~\ref{e2E_tsamp},
for stationary processes $\Cabcd(\qvec,\tsamp)$ and $\cabcd(\qvec,t)$
are closely related, cf.~Eq.~(\ref{eq_e2E_tsamp_Ctsamp}), and both correlation functions have the same
asymptote for large (sampling) times, cf.~Eq.~(\ref{eq_e2E_tsamp_Ctsamp_cinf}), 
i.e. the same {\em static} properties are probed.
From the numerical side $\Cabcd(\qvec,\tsamp)$ basically has the advantage that irrelevant high 
frequencies are filtered off (as it is usually the case for Einstein relations compared to 
Green-Kubo relations \cite{AllenTildesleyBook}). 
Compared with the common method to determine the system's inherent quenched stresses by 
quenching to the local energy minima \cite{Lemaitre15,Lemaitre18} this has the advantage that 
we measure the relevant stresses at the given finite temperature and not at a {\em different}
thermodynamic state. We believe that this method is therefore not only simpler (no additional quench 
required) but thermodynamically and conceptionally better defined. The technical downside is that 
time-series of many frames need to be stored.

\paragraph*{Symmetry breaking.}
The main physical reason for focusing on $t$-averaged stress fields $\Sab(\qvec,\tsamp)$ is, however, 
that the systematic time-averaging naturally allows to focus on the 
quenched stresses $\sabQ(\qvec)$ for each independent configuration $c$
by projecting out the trivial instantaneous thermal stress fluctuations
using that $\Sab(\qvec,\tsamp) \to \sabQ(\qvec)$ for large $\tsamp$. 
Importantly, the parameter $\Etilde$ is set by Eq.~(\ref{eq_ISFnormQ2Etilde}) in terms of the typical size 
of the quenched normal stresses $\ISFnormQ(\qvec)$ in reciprocal space using NRC. 
Due to Eq.~(\ref{eq_forceNRC}) these stresses may be finite without violating static mechanical 
equilibrium as shown by the finite attractor in Fig.~\ref{fig_sigrot_tsamp}. 

\paragraph*{Connection to thermodynamic response properties.}
Interestingly, $\Etilde$ was found to be numerically similar to the macroscopic 
Young modulus $E$ at thermal equilibrium in $d=2$.
As discussed in Sec.~\ref{e2E}, this finding was theoretically anticipated \cite{lyuda22a} 
since the predictions for equilibrium viscoelastic fluids,
must hold approximatively for our glasses due to the swap MC moves used for the system preparation.
(The same identification is obtained by an independent and complementary approach to long-wavelength
stress-correlations in glass-forming liquids based on the Zwanzig-Mori projection
operator formalism \cite{Fuchs17,Fuchs18,Fuchs19}.)
Altogether this implies
\begin{equation}
E = \Etilde \simeq \beta V \la \ISFnormQ(\qvec) \ISFnormQ(-\qvec) \ra
\label{eq_conc}
\end{equation}
for sufficiently small wavevectors.
The existence of a finite equilibrium Young modulus $E$ (and, hence, of a finite shear modulus $\mu$)
and the broken (continuous) symmetry of the stress field in reciprocal space,
characterized by the typical size of the quenched stresses $\ISFnormQ(\qvec)$,
are thus intimately connected.


\paragraph*{Outlook.}
The presented work suggests several natural extensions:
\begin{itemize}
\item 
The given tensor fields relations for isotropic systems naturally generalize to higher spatial dimensions, 
to tensor fields of different order and to crosscorrelation functions of different tensorial fields 
(say, between stress and strain fields) for which the major suffix symmetry Eq.~(\ref{eq_ten_sym_B}) may not hold
(which merely introduces one additional ICF).
\item
We have focused in the present work on Euclidean spaces and Carthesian coordinates.
It is possible to generalize our relations for systems embedded in non-Euclidean spaces,
say for glasses on spheres \cite{Tarjus15,Tarjus17}, 
and more general curvilinear coordinate systems \cite{McConnell,Lambourne}. 
Assuming the system to be {\em locally} isotropic the goal is always to construct from the 
available correlation functions the local (tangent) isotropic invariants and to check 
whether this can be done consistently with all available data sets.
\item
The proposed methodology could, e.g., also be used for irreversibly crosslinked polymer 
networks \cite{RubinsteinBook} or {\em active} non-equilibrium systems
\cite{Grosberg15,Xu20} which are both isotropic and highly viscous or even jammed. 
$\Etilde$ is in such cases just a fitting parameter characterizing the typical size
of the quenched stresses. 
\item
The presented work is also of relevance for the correlations of strain tensor fields 
\cite{Barrat14c,Reichman21,Reichman21b} and for the characterization of plastic deformations 
\cite{Voigtmann14,Fielding14,foot_ANS_comment}. 
It can be shown that the strain correlations of any isotropic elastic body are characterized by 
{\em two} ICFs which are, moreover, set by means of the general equipartition theorem \cite{ChaikinBook}
in the large-wavelength limit by the Lam\'e coefficients $\lambda$ and $\mu$.
This leads again to an octupolar correlation field pattern as seen in Fig.~\ref{fig_intro} for the stress.
\item
The low-$q$ limit $\Etilde(\tsamp)$ of $\beta V \Ctwo(q,\tsamp)$ 
becomes only a strictly $\tsamp$-independent constant for permanently quenched invariant 
stress fields $\ISFnormQ(\qvec)$, e.g., for permanent polymer networks \cite{RubinsteinBook}. 
By contrast, the quenched attractors $\ISFnormQ(\qvec)$ 
of low-temperature glasses become for $\tsamp$ similar to the
$\alpha$-relaxation time slow $\tsamp$-dependent fields $\ISFnormS(\qvec,\tsamp)$.
This is, e.g., of relevance for higher temperatures $T$ of the presented pLJ particle model 
where after a transient plateau $\Etilde(\tsamp)$ ultimately vanishes,
i.e. the long-range stress correlations disappear \cite{lyuda22a}. 
Future work should focus on the distributions of the $\ISFab(\qvec,\tsamp,T)$
as a direct means to systematically project out fast relaxation modes 
{\em independently} of the coordinate system.
\end{itemize}

\paragraph*{Acknowledgments.}
We are indebted to the HPC cluster of the University of Strasbourg for computational re\-sources.
 
\appendix

\section{Fourier transformation}
\label{FT}

\paragraph*{Continuous Fourier transform.}
Following Refs.~\cite{spmP4,lyuda22a} but at variance with Ref.~\cite{lyuda18}
we define in this work the Fourier transform (FT) 
$f(\qvec) = \Fcal[f(\rvec)]$ of a (real-valued) function $f(\rvec)$ in real space by
\begin{equation}
f(\qvec) = \frac{1}{V} \int \ddiff \rvec \ f(\rvec) \exp(-i \qvec \cdot \rvec)
\label{eq_FT_def}
\end{equation}
with $V$ being the volume of the system
and $\qvec$ a wavevector commensurate with the simulation box.
The inverse FT is then given by
\begin{equation}
f(\rvec) = \Fcal^{-1}[f(\qvec)] 
= \frac{V}{(2\pi)^d} \int \ddiff \qvec \ f(\qvec) \exp(i \qvec \cdot \rvec).
\label{eq_FT_def_inv}
\end{equation}
Note that $f(\rvec)$ and $f(\qvec)$ have the same dimension.
For notational simplicity the function names remain unchanged.
Let us denote by $f$ {\em without} argument the {\em macroscopic} field average.
Due to Eq.~(\ref{eq_FT_def}) we thus have the ``sum rule"
\begin{equation}
f(\qvec=\bfzero) = f \equiv \frac{1}{V} \int \ddiff \rvec \ f(\rvec).
\label{eq_FT_qzero}
\end{equation}
We note for later convenience the FTs
\begin{eqnarray}
\Fcal\left[\frac{\partial}{\partial  r_{\alpha}} f(\rvec)\right] & = & i q_{\alpha} f(\qvec) \label{eq_FT_A}\\
\Fcal\left[\delta(\rvec-\vvec)\right] & = & \frac{1}{V} \exp(-i \qvec \cdot \vvec) \label{eq_FT_B} \mbox{ and } \\
\Fcal\left[ \int_0^1 \ddiff s \ \delta(\rvec-\vvec s) \right]
& = & \frac{1}{V} \frac{1-\exp(-i \qvec \cdot \vvec)}{i \qvec \cdot \vvec}
\label{eq_FT_C}
\end{eqnarray}
with $\delta(\rvec)$ being Dirac's delta function.
Let us consider the spatial correlation function
\begin{equation}
c(\rvec) = \frac{1}{V} \int \ddiff \rvec' g(\rvec+\rvec') h(\rvec')
\label{eq_FT_correlation_r} 
\end{equation}
where the fields $g(\rvec)$ and $h(\rvec)$ are assumed to be real.
According to the ``correlation theorem" \cite{numrec} this becomes 
\begin{equation}
c(\qvec)  = g(\qvec) h^{\star}(\qvec) =  g(\qvec) h(-\qvec)
\label{eq_FT_correlation_q} 
\end{equation}
in reciprocal space (with $\star$ marking the complex conjugate).
For auto-correlation functions, i.e. for $g(\rvec)=h(\rvec)$,
this simplifies to (``Wiener-Khinchin theorem") 
\begin{equation}
c(\qvec)  = g(\qvec) g^{\star}(\qvec) = |g(\qvec)|^2,
\label{eq_FT_WKT} 
\end{equation}
i.e. the Fourier transformed auto-correlation functions are real and $\ge 0$ for all $\qvec$.
Moreover, we shall consider correlation functions $c(\rvec)$, 
Eq.~(\ref{eq_FT_correlation_r}), being {\em even} in real space, $c(\rvec)=c(-\rvec)$, 
and thus also in reciprocal space, $c(\qvec)=c(-\qvec)=c^{\star}(\qvec)$,
i.e. $c(\qvec)$ is real. 

For many reasons it is convenient to consider correlation functions $c(\rvec)$
which vanish for large $r=|\rvec|$. This is achieved here 
by replacing in Eq.~(\ref{eq_FT_correlation_r}) 
$g(\rvec)$ by $\tilde{g}(\rvec)=g(\rvec)-g$ and 
$h(\rvec)$ by $\tilde{h}(\rvec)=h(\rvec)-h$ with $g$ and $h$ being field averages
according to Eq.~(\ref{eq_FT_qzero}). We thus probe in this work correlation functions
\begin{eqnarray}
c(\rvec) & = & \frac{1}{V} \int \ddiff \rvec' \tilde{g}(\rvec+\rvec') \tilde{h}(\rvec')
\nonumber \\
& = & \frac{1}{V} \int \ddiff \rvec' g(\rvec+\rvec') h(\rvec') - g h
\label{eq_FT_correlation_r_reference} 
\end{eqnarray} 
using a real offset $g h$. Eq.~(\ref{eq_FT_correlation_q}) thus becomes
\begin{equation}
c(\qvec) = \tilde{g}(\qvec) \tilde{h}(-\qvec) 
= \left\{\begin{array}{ll}
g(\qvec) h(-\qvec)  & \mbox{ for } \qvec \ne \bfzero \\
0                   & \mbox{ for } \qvec = \bfzero
\end{array}\right.
\label{eq_FT_correlation_q_reference} 
\end{equation}
where we have used that $g=g(\qvec=\bfzero)$ and $h=h(\qvec=\bfzero)$.

\paragraph*{Discrete FT on microcell grid.}
Numerically, all fields $f(\rvec)$ are stored on a regular equidistant grid 
as shown in Fig.~\ref{fig_grid}. 
Periodic boundary conditions are assumed \cite{AllenTildesleyBook}. 
The discrete FT and its inverse become
\begin{eqnarray}
f(\qvec) & = & \frac{1}{\nVgrid} \sum_{\rvec} f(\rvec) \exp(-i \qvec\cdot \rvec)
\label{eq_FT_discrete} \\
f(\rvec) & = & \sum_{\qvec} f(\qvec) \exp(i \qvec\cdot \rvec)
\label{eq_FT_discrete_inv} 
\end{eqnarray}
with $\sum_{\rvec}$ and $\sum_{\qvec}$ being discrete sums over $\nVgrid=V/\agrid^d$ grid points in,
respectively, real or reciprocal space. With $\nLgrid$ being the number of grid points
in each spatial direction $\alpha$, i.e. $\nVgrid=\nLgrid^d$ and $L=\nLgrid \agrid$, we have
\begin{equation}
\frac{r_{\alpha}}{\agrid} = n_{\alpha} \mbox{ and } q_{\alpha} \agrid = \frac{2\pi}{\nLgrid} n_{\alpha}
\label{eq_FT_gridpoints}
\end{equation}
with integers $n_{\alpha}=0,1,\ldots, \nLgrid-1$.
To take advantage of the implemented Fast-Fourier transform (FFT) routines 
\cite{numrec} $\nLgrid$ is an integer-power of $2$.
The real space correlation function $c(\rvec)$ on the discrete grid is 
\begin{equation}
c(\rvec) = \frac{1}{\nVgrid} \sum_{\rvec'} g(\rvec+\rvec') h(\rvec') - g h
\label{eq_FT_correlation_r_grid}
\end{equation}
being an operation of order $\Ocal(\nVgrid^2)$.
Importantly, the periodicity of all fields must be taken into account.
It is obviously much more efficient to first FFT the discrete fields $g(\rvec)$ and $h(\rvec)$
or $\tilde{g}(\rvec)$ and $\tilde{h}(\rvec)$ 
and to apply then Eq.~(\ref{eq_FT_correlation_q}) \cite{AllenTildesleyBook,numrec}.
In this manner $c(\qvec)$ is automatically {\em periodic} in all spatial directions
of the simulation box and the same applies to the correlation function
$c(\rvec) = \Fcal^{-1}[c(\qvec)]$ in real space computed using Eq.~(\ref{eq_FT_discrete_inv}).

\section{Fourier transforms in two dimensions}
\label{q2r}

\paragraph*{Correlation functions in reciprocal space.}
As shown in Fig.~\ref{fig_Crot_q} 
all four ICFs in $d=2$ become constant in reciprocal space for sufficiently small
wavevectors $q$ and large sampling times $\tsamp$, but only
$\beta V \Ctwo(q) \simeq \Etilde$ remains finite. 
According to Eq.~(\ref{eq_ten_Tabcd_d2_BB}) of Sec.~\ref{ten_Tabcd_d2}
we thus have
\begin{eqnarray}
\beta V \Cabcd(\qvec) & \simeq & \Etilde \times \label{eq_q2r_start} \\
(\delta_{\alpha\beta}\delta_{\gamma\delta} & - &
\qhat_{\alpha} \qhat_{\beta} \delta_{\gamma\delta} -
\qhat_{\gamma} \qhat_{\delta} \delta_{\alpha\beta} +
\qhat_{\alpha} \qhat_{\beta} \qhat_{\gamma} \qhat_{\delta})
\nonumber
\end{eqnarray}
for the correlation functions in the ``old" (unrotated) reference frame.
Hence, we have, e.g.,
\begin{eqnarray}
\hspace*{-0.5cm}\beta V \Cxyxy(\qvec) 
& = & \nonumber \\ 
\beta V \Cxxyy(\qvec) & \simeq & \Etilde \ \qhat_1^2 \qhat_2^2 \label{eq_q2r_Cxyxy_qq}\\
\beta V \Cxxxx(\qvec) & \simeq & \Etilde \left(1- 2\qhat_1^2 + \qhat_1^4\right) 
= \Etilde \qhat_2^4 \label{eq_q2r_Cxxxx_qq} \\
\beta V \Cyyyy(\qvec) & \simeq & \Etilde \left(1- 2\qhat_2^2 + \qhat_2^4\right) 
= \Etilde \qhat_1^4 \label{eq_q2r_Cyyyy_qq} \\
\beta V \Cxxxy(\qvec) & \simeq & -\Etilde \ \qhat_1\qhat_2^3 
\label{eq_q2r_Cxxxy_qq} \\
\beta V \Cyyyx(\qvec) & \simeq & -\Etilde \ \qhat_1^3\qhat_2.
\label{eq_q2r_Cyyyx_qq}
\end{eqnarray}
Let us rewrite these cases in terms of the angle $\thetaq$ of the normalized wavevector 
$\qhatvec = \ (\cos(\thetaq),\sin(\thetaq))$.
Using standard trigonometric relations it is readily seen that
\begin{eqnarray}
- \beta V \Cxyxy(\qvec) & = & \nonumber \\
- \beta V \Cxxyy(\qvec) & \simeq & 
\frac{\Etilde}{8} \left(\cos(4 \thetaq)-1\right) \label{eq_q2r_Cxyxy_q} \\
\beta V
(\Cxxxx(\qvec)+\Cyyyy(\qvec))/2 & \simeq & 
\frac{\Etilde}{8} \left(\cos(4 \thetaq)+3\right) \label{eq_q2r_Cxxxx_mean_q} \\
\beta V
(\Cxxxx(\qvec)-\Cyyyy(\qvec))/2 & \simeq & -\frac{\Etilde}{2} \cos(2\thetaq) \label{eq_q2r_Cxxxx_diff_q}\\
\beta V
(\Cxxxy(\qvec)+\Cyyyx(\qvec))/2 & \simeq & -\frac{\Etilde}{4} \sin(2\thetaq) \label{eq_q2r_Cxxxy_mean_q}\\
\beta V 
(\Cxxxy(\qvec)-\Cyyyx(\qvec))/2 & \simeq & \frac{\Etilde}{8} \sin(4\thetaq) \label{eq_q2r_Cxxxy_diff_q}
\end{eqnarray}
in the above-mentioned limits. More generally, all correlation functions in 
two dimensions can be expressed by a linear superposition of the orthogonal basis functions 
$\cos(p \thetaq)$ and $\sin(p \thetaq)$ with $p=0$, $2$ and $4$.

\paragraph*{Inverse Fourier transform.}
To obtain the correlation functions in real space we thus have to compute
the inverse FT in $d=2$ dimensions for
\begin{equation}
f(\qvec) = \frac{1}{V} \  v(q) \ b_{p}(\thetaq)
\label{eq_q2r_vqwpq}
\end{equation}
with $b_{p}(\thetaq)$ standing for the basis functions $\cos(p \thetaq)$ or $\sin(p \thetaq)$
with $p=2$ and $4$. The constant terms ($p=0$) in Eq.~(\ref{eq_q2r_Cxyxy_q}) and 
Eq.~(\ref{eq_q2r_Cxxxx_mean_q})
are irrelevant for our considerations leading merely to $\delta$-contributions at $r=0$.
The yet unspecified auxiliary function $v(q)$, with $v(q) \to \Etilde$ for $q \to 0$,
will be chosen below in a convenient manner to take 
advantage of special mathematical functions \cite{abramowitz}.
Hence, using Eq.~(\ref{eq_FT_def_inv}) we have
\begin{eqnarray}
f(\rvec) 
 & = & \frac{1}{4\pi^2} \int \ddiff q \ q v(q) \times \nonumber \\
 & & \hspace*{-.5cm} \int_0^{2\pi} \ddiff \thetaq \ b_{p}(\thetaq) \exp[i q r \cos(\thetaq-\thetar)]
\label{eq_q2r_invFT}
\end{eqnarray}
with $\thetar$ being the angle of $\rhatvec = (\cos(\thetar),\sin(\thetar))$.
We make now the substitution $\theta = \thetaq-\thetar$ and use that \cite{abramowitz}
\begin{eqnarray}
\cos(p\theta+p\thetar)+\cos(-p\theta+p\thetar) & = & 2 \cos(p\theta) \cos(p\thetar) 
\nonumber \\ 
\sin(p\theta+p\thetar)+\sin(-p\theta+p\thetar) & = & 2 \cos(p\theta) \sin(p\thetar).
\nonumber
\end{eqnarray}
We remind that following Eq.~(9.1.21) of Ref.~\cite{abramowitz} 
the integer Bessel function $J_p(z)$ may be written
\begin{equation}
J_p(z) = \frac{i^{-p}}{\pi} \int_0^{\pi} \ddiff \theta \cos(p\theta) \exp[i z \cos(\theta)].
\label{eq_q2r_Jp}
\end{equation}
This leads to
\begin{equation}
f(\rvec) = \frac{i^p}{2\pi} b_{p}(\thetar) \ \int \ddiff q \ q v(q) \ J_p(r q). 
\label{eq_q2r_JpB}
\end{equation}
We use next Eq.~(11.4.28) of Ref.~\cite{abramowitz} 
\begin{eqnarray}
\int_0^{\infty} e^{-(at)^2} t^{\mu-1} J_{\nu}(bt) \ddiff t
& = &
\frac{\Gamma[(\nu+\mu)/2] \ (b/2a)^{\nu}}{2a^{\mu}\Gamma(\nu+1)} \nonumber \\
& & \hspace*{-2.5cm}\times \ M[(\nu+\mu)/2,\nu+1,-(b/2a)^2]
\label{eq_q2r_J2M}
\end{eqnarray}
for $\Re(\mu+\nu)>0$ and $\Re(a^2)>0$ relating the general Bessel function $J_{\nu}(b t)$ to 
the confluent hypergeometric Kummer function $M(a,b,z)$.
($\Gamma(x)$ denotes the standard Gamma function \cite{abramowitz}.)
To take advantage of Eq.~(\ref{eq_q2r_J2M}) we finally set 
$v(q) = \Etilde \exp[-(aq)^2]$.
Note that $v(q) \to \Etilde$ for $a q \to 0$, i.e. the auxiliary variable $a$ becomes irrelevant for small $q$. 
We thus rewrite Eq.~(\ref{eq_q2r_JpB}) as
\begin{eqnarray}
f(\rvec) & = &
\Etilde \ \frac{i^p}{2 \pi} \frac{\Gamma(p/2+1) (r/2a)^p}{2a^2\Gamma(p+1)}
\ \ b_{p}(\thetar)
 \nonumber \\
& \times & M[p/2+1,p+1,-(r/2a)^2] 
\label{eq_q2r_M} 
\end{eqnarray}
in terms of Kummer's function $M$. Following Eq.~(13.1.5) of Ref.~\cite{abramowitz}
\begin{equation}
M(a,b,z) = \frac{\Gamma(b)}{\Gamma(b-a)} (-z)^{-a} \left( 1 + \Ocal(|z|^{-1} \right)
\label{eq_Ft_M2}
\end{equation}
for a real part $\Re(z) < 0$. Using this expansion in Eq.~(\ref{eq_q2r_M})
finally leads to
\begin{equation}
f(\rvec) \simeq \Etilde \ \frac{i^p p}{2\pi r^2} \ b_{p}(\thetar) 
\label{eq_q2r_fr_p}
\end{equation}
for sufficiently large $r$ and $p >0$.
We note that the auxiliary variable $a$ indeed drops out,
that $f(\rvec)=f(-\rvec)$ and that $f(\rvec)$ is real for even $p$. 
 
\paragraph*{Correlation functions in real space.}
Using Eq.~(\ref{eq_q2r_fr_p}) we can finally restate 
Eqs.~(\ref{eq_q2r_Cxyxy_q}-\ref{eq_q2r_Cxxxy_diff_q})
in real space as
\begin{eqnarray}
-\beta \Cxxyy(\rvec) =
-\beta \Cxyxy(\rvec) & = & \nonumber \\
\beta \left(\Cxxxx(\rvec) + \Cyyyy(\rvec)\right)/2 
  & \simeq & \frac{\Etilde}{4\pi r^2} \cos(4 \thetar) \label{eq_q2r_Cxxxx_mean_r} \\
\beta \left(\Cxxxx(\rvec) - \Cyyyy(\rvec)\right)/2 
  & \simeq & \frac{\Etilde}{2\pi r^2} \cos(2 \thetar) \label{eq_q2r_Cxxxx_diff_r} \\
\beta \left(\Cxxxy(\rvec) + \Cyyyx(\rvec)\right)/2 & \simeq & \frac{\Etilde}{4\pi r^2} \sin(2 \thetar) 
\label{eq_q2r_Cxxxy_mean} \\
\beta \left(\Cxxxy(\rvec) - \Cyyyx(\rvec)\right)/2 & \simeq & 
\frac{\Etilde}{4\pi r^2} \sin(4 \thetar) \label{eq_q2r_Cxxxy_diff_r}
\end{eqnarray}
for $r > 0$ and $\tsamp \to \infty$.
Some of these relations are shown in Fig.~\ref{fig_Cabcd_thetar} and Fig.~\ref{fig_Cabcd_rproj} 
in the main part of this work.
 
\section{Additional computational details}
\label{comp}

\paragraph*{Simulation model.}
We consider systems of polydisperse Lennard-Jones (pLJ) particles in $d=2$ dimensions
where two particles $i$ and $j$ of diameter $D_i$ and $D_j$ interact by means of a
central pair potential \cite{WXP13,lyuda19a,spmP1,spmP2,lyuda21b,lyuda22a} 
\begin{equation}
u(s) = 4 \epsilon \left(\frac{1}{s^{12}}-\frac{1}{s^{6}}\right)
\mbox{ with } s=\frac{r}{(D_i+D_j)/2}
\label{eq_comp_Us}
\end{equation}
being the reduced distance according to the Lorentz rule \cite{HansenBook}.
This potential is truncated and shifted \cite{AllenTildesleyBook}
with a cutoff $\scut=2\smin$ given by the minimum $\smin$ of $u(s)$.
Lennard-Jones units \cite{AllenTildesleyBook} are used throughout this study,
i.e. $\epsilon=1$ and the average particle diameter is set to unity.
The diameters are uniformly distributed between $0.8$ and $1.2$.
We also set Boltzmann's constant $\kB=1$ and assume that all particles have the same mass $m=1$. 
The last point is irrelevant for the presented Monte Carlo (MC) simulations
\cite{AllenTildesleyBook}.
Time is measured in units of MC cycles throughout this work.

\paragraph*{Operational parameters.}
We focus in the present work on configurations with $n=10000$ particles
albeit we have sampled a broad range of particle numbers $n$ between $n=100$ and $n=160000$
(cf. inset of Fig.~\ref{fig_Cabcd_rproj}).
$\Nc=200$ independent configurations $c$ have been obtained by quenching 
configurations equilibrated at a high temperature $T=0.55$ in the liquid limit.
This is done using a combination of local MC moves 
\cite{AllenTildesleyBook} and swap MC moves
exchanging the sizes of pairs of particles \cite{Berthier17,spmP1}.
In addition an MC barostat \cite{AllenTildesleyBook} imposes an average 
normal stress $P=2$ \cite{WXP13,spmP1}. 

\paragraph*{Working temperature.}
At the working temperature $T=0.2$ we first thoroughly temper over $\tsamp=10^7$ 
all configurations with switched-on local, swap and barostat MC moves
and then again over $\tsamp=10^7$ with switched-on local and swap moves and switched-off barostat moves.
The shear-stress relaxation function $G(\tau)$ \cite{WXB15,WXBB15,spmP1,lyuda22a}
for this tempering run is shown in Fig.~\ref{fig_Gt}.
As can be seen (squares), $G(\tau)$ rapidly decays to a shear modulus $\mu=0$ (dashed line)
as expected for an equilibrium liquid.
The final production runs are carried out at constant volume $V$ only keeping the slow local moves.
Importantly, $T=0.2$ is well below the known glass transition temperature 
$\Tglass \approx 0.26$ assuming only local MC hopping moves \cite{WXP13,spmP1}.
As shown by the circles in Fig.~\ref{fig_Gt} only keeping the local moves
essentially traps the (equilibrated) configurations in local metabasins
where they behave as elastic bodies with a finite shear modulus $\mu \approx 14$ (solid line).
Due to the barostat used for the quenching of the configuration
the box volume $V=L^d$ differs slightly between different configurations $c$
while $V$ is identical for all (correlated) configurations $t$ of the time-series
of the same independent configuration $c$.
In all cases the number density is of order unity, 
i.e. the particle number $n$ and the volume $V$ are similar.
This implies that the ideal pressure $\Pid = T n/V \approx 0.2$ is much smaller
than the imposed total pressure $P=\Pid+\Pex=2$. 
As shown elsewhere \cite{lyuda19a,spmP1,spmP2,lyuda21b,lyuda22a}
our systems are homogeneous and isotropic and crystallization is irrelevant.

\paragraph*{Data sampling.}
For each of the $\Nc=200$ independent configurations $c$ 
we sample and store {\em four} ensembles of time series containing each $\Nt=10000$ 
instantaneous ``frames" $t$.
These are obtained using the equidistant time intervals $\tincr=1$, $10$, $100$ and $1000$
\cite{foot_t_tau}. 
This implies for each time series a largest sampling time 
$\tsamp = \Nt \tincr$, i.e. $\tsamp=10^7$ for the largest time interval. 
This was chiefly done to check that all correlation functions $\Cabcd(\rvec)$ become indeed
$\tsamp$-independent as shown in Fig.~\ref{fig_sigrot_tsamp} and Fig.~\ref{fig_ICFs_tlong}. 
The time-averaged stress tensors $\bar{\sigma}_{\alpha\beta}|_c$ 
and stress tensor fields $\bar{\sigma}_{\alpha\beta}(\rvec)|_c$
are obtained by averaging over
the corresponding instantaneous $\sigma_{\alpha\beta}(t)|_c$ 
and $\sigma_{\alpha\beta}(\rvec,t)|_c$ using each frame $t$ of a time-series.
We thus obtain first 
\begin{equation}
\left.\Cabcd(\qvec)\right|_c=\left\{ 
\begin{array}{ll}
\left.\bar{\sigma}_{\alpha\beta}(\qvec)\right|_c \left.\bar{\sigma}_{\gamma\delta}(-\qvec)\right|_c
& \mbox{ for } \qvec \ne \bfzero \\
0                   & \mbox{ for } \qvec = \bfzero
\end{array}\right.
\label{eq_comp_Cabcd_c}
\end{equation}
for each $c$ according to Eq.~(\ref{eq_FT_correlation_q_reference}), take then the $c$-average
\begin{equation}
\Cabcd(\qvec)= \frac{1}{\Nc} \sum_{c=1}^{\Nc} \left.\Cabcd(\qvec)\right|_c,
\label{eq_comp_caver_Cabcd}
\end{equation}
and perform finally the inverse FFT to real space
\begin{equation}
\Cabcd(\rvec)=\Fcal^{-1}[\Cabcd(\qvec)].
\end{equation}

\section{Construction of stress field}
\label{stress}

\paragraph*{Macroscopic stresses.}
Carets ``$\hat{a}$" mark here instantaneous properties and the argument $t$ is dropped.
Let us first remind the Irving-Kirkwood formula for the total macroscopic ($\qvec=\bfzero$) 
stress tensor $\sighatab$ \cite{IrvingKirkwood50,AllenTildesleyBook}. 
Note that the total system Hamiltonian is the sum of an ideal contribution, 
depending only on the momenta, and an excess contribution, depending only on the particle positions. 
This implies \cite{AllenTildesleyBook} that the total stress tensor $\sighatab=\sighatidab+\sighatexab$
is a sum of an ideal stress $\sighatidab$ and an excess stress $\sighatexab$.
The ideal contribution is \cite{AllenTildesleyBook}
\begin{equation}
\sighatidab = -\frac{1}{V} \sum_{a=1}^n \frac{p^a_{\alpha}p^a_{\beta}}{m}
\label{eq_stress_sigid_tot}
\end{equation}
with $p^a_{\alpha}$ being the $\alpha$-component of the momentum of particle $a$.
This contribution is naturally not accessible in an MC simulation.
Since we want anyway ultimately to time-average all instantaneous stresses,
Eq.~(\ref{eq_stress_sigid_tot}) may be replaced without loss of information
by its ensemble average $\langle \sighatidab \rangle = - \Pid \delta_{\alpha\beta}$ 
with $\Pid=T n/V$ \cite{AllenTildesleyBook}. 
The excess contribution is given by the sum $\sighatexab = \sum_l w^l_{\alpha\beta}/V$
over all interacting pairs $l$ of particles $a < b$.
Let us denote by $\rvec^l=\rvec^b-\rvec^a$ the vector from the position $\rvec^a$ of $a$
to the position $\rvec^b$ of $b$, by $r^l = |\rvec^l|$ its length and by $\hat{r}^l_{\alpha}$ 
a component of its unit vector. The contribution $w^l_{\alpha\beta}$
of the interaction $l$ to the excess stress is then \cite{AllenTildesleyBook}
\begin{equation}
w^l_{\alpha\beta} \equiv r_l u^{\prime}(r_l) \hat{r}^l_{\alpha} \hat{r}^l_{\beta}
\label{eq_stress_wl}
\end{equation}
with $u^{\prime}(r)$ being the first derivative of the pair potential. 

\paragraph*{Stress tensor fields in reciprocal space.}
The corresponding stress tensor field 
$\sighatab(\qvec) = \sighatidab(\qvec)+\sighatexab(\qvec)$ in reciprocal space
may be directly obtained from the local momentum equation 
\cite{LandauElasticity,HansenBook}
\begin{equation}
\frac{\partial}{\partial t} \hat{P}_{\alpha}(\rvec) = 
\frac{\partial}{\partial r_{\beta}} \hat{\sigma}_{\alpha\beta}(\rvec)
\label{eq_stress_momentum_r}
\end{equation}
as shown long ago \cite{Lutsko88,HansenBook}.
Note that the momentum density $\hat{P}_{\alpha}(\rvec)$ is defined by \cite{HansenBook}
\begin{equation}
\hat{P}_{\alpha}(\rvec) = \sum_{a=1}^n p^{a}_{\alpha} \delta(\rvec-\rvec^{a})
\label{eq_stress_momentum_density_r}
\end{equation}
in terms of the individual particle momenta and positions.
Rewritten in Fourier space Eq.~(\ref{eq_stress_momentum_r}) becomes
\begin{eqnarray}
i q_{\beta} \hat{\sigma}_{\alpha\beta}(\qvec)
&= &  \frac{\partial}{\partial t} \hat{P}_{\alpha}(\qvec) \label{eq_stress_momentum_start} \\
& = & -i q_{\beta}\frac{1}{V} \sum_{a=1}^n \frac{p^a_{\alpha} p^a_{\beta}}{m} \exp(-i\qvec\cdot\rvec^{a})
\nonumber \\
& + & \frac{1}{V} \sum_{a=1}^n \dot{p}^{a}_{\alpha} \exp(-i\qvec\cdot\rvec^{a}).
\label{eq_stress_momentum_q_B}
\end{eqnarray}
The first term in the previous relation gives the ideal stress contribution \cite{Lutsko88,HansenBook}
\begin{equation}
\sighatidab(\qvec) = -\frac{1}{V} \sum_{a=1}^n \frac{p^a_{\alpha} p^a_{\beta}}{m} 
\exp(-i\qvec \cdot \rvec^a).
\label{eq_stress_sigid_q}
\end{equation}
As before for the macroscopic ideal stress we may integrate out the momenta $\{\pvec\}$
and replace Eq.~(\ref{eq_stress_sigid_q}) by 
\begin{equation}
\sighatidab(\qvec) = -\frac{T}{V} \delta_{\alpha\beta} \sum_{a=1}^n \exp(-i\qvec \cdot \rvec^a)
\label{eq_stress_sigid_q_MC}
\end{equation}
which is used for our MC simulations.
Using Newton's second law for $\dot{p}^{a}_{\alpha}$ and a total potential energy given 
by pair interactions such as Eq.~(\ref{eq_comp_Us}) the excess stress in reciprocal space is 
obtained from the second term in Eq.~(\ref{eq_stress_momentum_q_B}) as shown 
in Refs.~\cite{Lutsko88,HansenBook}. This gives 
\begin{equation}
\sighatexab(\qvec)  = \frac{1}{V} \sum_l w^l_{\alpha\beta} 
\frac{e^{-i\qvec \cdot \rvec^b}-e^{-i\qvec\cdot \rvec^a}}{-i \qvec\cdot (\rvec^b-\rvec^a)}.
\label{eq_stress_sigex_q_A}
\end{equation}
Let us define $\mvec^l=(\rvec^a+\rvec^b)/2$ and $\hvec^l=(\rvec^b-\rvec^a)/2$.
As noted by Lema\^\i tre \cite{Lemaitre14} the previous relation may be rewritten
more conveniently as
\begin{equation}
\sighatexab(\qvec) = \frac{1}{V} \sum_l w^l_{\alpha\beta} 
\frac{\sin(\qvec\cdot \hvec^l)}{\qvec\cdot \hvec^l}
e^{-i \qvec\cdot \mvec^l}.
\label{eq_stress_sigex_q}
\end{equation}
We remark first that $\sighatidab(\qvec)$ and $\sighatexab(\qvec)$ reduce properly for $\qvec \to \bfzero$
to the macroscopic stress contributions noted above.
Note also that the term $\sin(x)/x \to 1$ for $x=\qvec\cdot \hvec^l \ll 1$, 
i.e. for wavelengths $2\pi/q$ larger than the typical interaction range of order unity 
in the present model. In this limit Eq.~(\ref{eq_stress_sigex_q}) further simplifies to
\begin{equation}
\sighatexab(\qvec) \approx \frac{1}{V} \sum_l w^l_{\alpha\beta} e^{-i \qvec\cdot \mvec^l},
\label{eq_stress_sigex_q_B}
\end{equation}
i.e. only the mean position $\mvec^l$ of two particles matters, 
not their relative orientation. Since this study anyway focuses on the universal large-wavelength limit,
the approximation Eq.~(\ref{eq_stress_sigex_q_B}) should be sufficient.

\paragraph*{Computation in reciprocal space.}
Using Eq.~(\ref{eq_stress_sigid_q_MC}) and Eq.~(\ref{eq_stress_sigex_q}) or 
Eq.~(\ref{eq_stress_sigex_q_B}) one may directly compute $\sighatab(\qvec)$
for a given configuration using a discrete grid of linear length $\nLgrid$
as shown in Fig.~\ref{fig_grid}. Since we anyway need the stress fields in reciprocal 
space to obtain the correlation functions, Eq.~(\ref{eq_FT_correlation_q_reference}), 
this is the most direct method and a useful exercise to test less direct definitions and approximations.
Unfortunately, this is numerically not the most efficient procedure since it evolves 
$\nLgrid^d \times n \simeq V^2$ operations. It is computationally much faster to first 
obtain the stress fields in real space with a number of operations of order $n \propto V$ 
and then to FFT transform with a number of operations of order $V \log(V)$.

\paragraph*{Stress tensor fields in real space.}
We thus need to state the corresponding relations in real space.
Inverse Fourier transformation yields for Eq.~(\ref{eq_stress_sigid_q}) 
\begin{equation}
\sighatidab(\rvec) = -\sum_{a=1}^n \frac{p^a_{\alpha} p^a_{\beta}}{m}
\ \delta(\rvec-\rvec^a) \label{eq_stress_sigid_r} 
\end{equation}
and for the preaveraged ideal stress, Eq.~(\ref{eq_stress_sigid_q_MC}), 
\begin{equation}
\sighatidab(\rvec) = -T \delta_{\alpha\beta} \sum_{a=1}^n  \delta(\rvec-\rvec^a).
\label{eq_stress_sigid_r_MC}
\end{equation}
One confirms using Eq.~(\ref{eq_FT_C}) that Eq.~(\ref{eq_stress_sigex_q_A}) becomes 
\begin{eqnarray}
\sighatexab(\rvec) & = &  \sum_l w_{\alpha\beta}^l I_l(\rvec) \mbox{ with } 
\label{eq_stress_sigex_r} \\
I_l(\rvec) & = & \int_0^1 \ddiff s \ \delta(\rvec-(\rvec^a+\rvec^ls)) \nonumber 
\end{eqnarray}
being a line integral between the two particle positions $\rvec^a$ and $\rvec^b$ of the interaction $l$.
This relation has an old history going at least back to the work by Kirkwood and Buff
\cite{BuffKirkwood49} for the layer-resolved slabs of microcells \cite{AllenTildesleyBook}.
Various rediscoveries, reformulations and generalizations (e.g., for multibody potentials 
and constraint dynamics) of the inverse Fourier transform 
Eq.~(\ref{eq_stress_sigex_r}) are discussed elsewhere 
\cite{TadmorMMBook,BuffKirkwood49,IrvingKirkwood50,Noll55,Varnik00,Goldhirsch02,VanegasArroyo14,Lemaitre14,MM21}.
If we use instead the approximation Eq.~(\ref{eq_stress_sigex_q_B}) the line integral
is replaced by
\begin{equation}
I_l(\rvec) \approx \delta(\rvec-\mvec^l). \label{eq_stress_sigex_r_Iapprox}
\end{equation}
We emphasize that this coarse-grained expression is completely sufficient for most applications.
(The $\delta$-functions in Eq.~(\ref{eq_stress_sigex_r}) or Eq.~(\ref{eq_stress_sigex_q_B})
are sometimes ``blurred" using more general weighting distributions \cite{Lemaitre14}.)
Please note that for all stated expressions the volume averages are identical to the
macroscopic relation, i.e.
\begin{eqnarray}
\sighatidab & = & \frac{1}{V} \int \ddiff \rvec \ \sighatidab(\rvec) = \sighatidab(\qvec=\bfzero)
\label{eq_stress_sigid_r_sumrule} \\
\sighatexab & = & \frac{1}{V} \int \ddiff \rvec \ \sighatexab(\rvec) = \sighatexab(\qvec=\bfzero)
\label{eq_stress_sigex_r_sumrule} 
\end{eqnarray}
hold exactly without approximation.

\begin{figure}[t]
\centerline{\resizebox{1.0\columnwidth}{!}{\includegraphics*{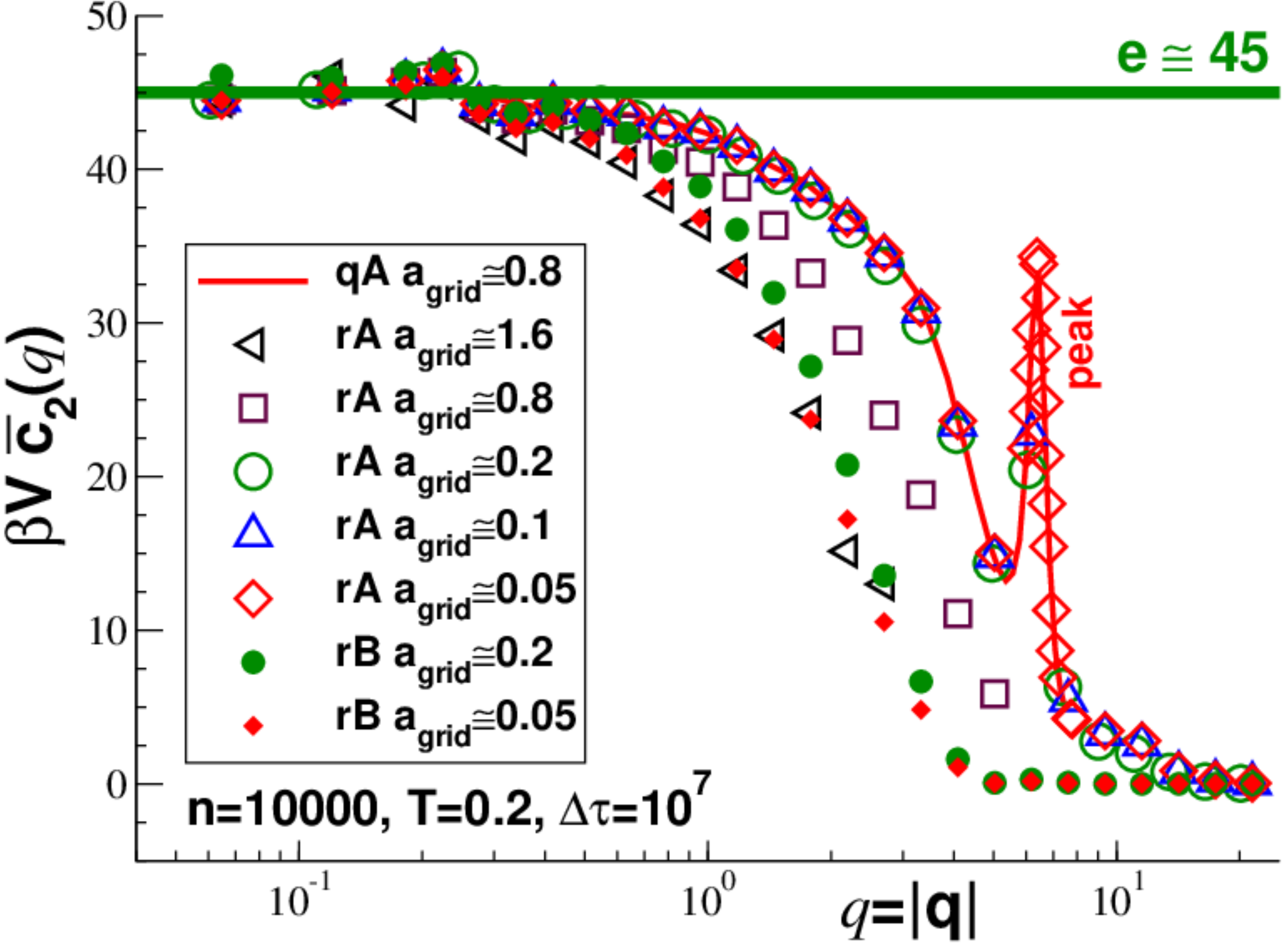}}}
\caption{Rescaled ICF $\beta V \Ctwo(q)$
comparing different methods and $\agrid=L/\nLgrid$.
The method ``qA" uses Eq.~(\ref{eq_stress_sigex_q}) directly in reciprocal space
(solid line),
the method ``rA" the line integration Eq.~(\ref{eq_stress_sigex_r}) in real space
(open symbols),
the method ``rB" the coarse-graining approximation
Eq.~(\ref{eq_stress_sigex_r_Iapprox}) in real space (filled symbols).
The same result is obtained in all cases for sufficiently small $q$.
The position of the peak of $\beta V \Ctwo(q)$ at $q \approx 6.5$ is 
very similar to the position of the main peak of the structure factor $S(q)$
as shown in Ref.~\cite{lyuda21b}.}
\label{fig_stress_Ctwo_q}
\end{figure}

\paragraph*{Discrete grid implementations.}
The numerical calculation of the above continuous space relations on a discrete grid of 
linear length $\nLgrid$ (cf. Fig.~\ref{fig_grid}) is obvious for the ideal contributions, 
Eq.~(\ref{eq_stress_sigid_q_MC}) and Eq.~(\ref{eq_stress_sigid_r_MC}), and also 
for the excess contribution Eq.~(\ref{eq_stress_sigex_q}) in reciprocal space or the 
coarse-graining approximations Eq.~(\ref{eq_stress_sigex_q_B}) or 
Eq.~(\ref{eq_stress_sigex_r_Iapprox}).
Slightly less trivial is the implementation of the line integral Eq.~(\ref{eq_stress_sigex_r}).
Different variants exist for distributing the information of an interaction $l$ 
on the grid as discussed in the literature 
\cite{AllenTildesleyBook,TadmorMMBook}.
In our view this issue is not crucial since the corresponding small wavelengths
have no universal physical meaning being due to an artificial computer model and
are, moreover, readily renormalized away as shown by Eq.~(\ref{eq_stress_sigex_q_B}). 
An important technical point is only that any reasonable method must strictly obey 
Eq.~(\ref{eq_stress_sigex_r_sumrule}), i.e. that the contributions to all grid
points from the line integral must be properly weighted. We use a simple
numerical rendering of Eq.~(\ref{eq_stress_sigex_r}) on the grid:
for each of $k$ equidistant points on the continuous line between $\rvec^a$ and $\rvec^b$
the closest grid point is incremented by $w_{\alpha\beta}^l/k$.
We do not care if sometimes a grid point gets several or even all contributions
(which happens if $\agrid$ is large) or none (which happens especially if the grid is too fine). 
All data presented in other parts of this work have been obtained using Eq.~(\ref{eq_stress_sigid_r})
and Eq.~(\ref{eq_stress_sigex_r}) with $k=20$ and a grid spacing $\agrid \approx 0.2$.
Different variants are compared in Fig.~\ref{fig_stress_Ctwo_q} 
where we focus on the rescaled ICF $\beta V \Ctwo(q)$.
We compare results obtained with
\begin{itemize}
\item
method ``qA" using Eq.~(\ref{eq_stress_sigid_q_MC}) and Eq.~(\ref{eq_stress_sigex_q}),
\item
method ``rA" using Eq.~(\ref{eq_stress_sigid_r_MC}) and Eq.~(\ref{eq_stress_sigex_r}),
\item
method ``rB" using Eq.~(\ref{eq_stress_sigid_r_MC}) and Eq.~(\ref{eq_stress_sigex_r_Iapprox})
\end{itemize}
for different grid constants $\agrid=L/\nLgrid$ as indicated.
Method qA ($\nLgrid=128$, solid line) required a month of computation on a local workstation 
cluster with 64 nodes while all other examples were computed within a couple of hours on
the same cluster. 
Most importantly, all methods are seen to yield similar results for small $q$ and 
the differences are minor for wavevectors up to $q \approx 1$.
The observed differences between the exact relation Eq.~(\ref{eq_stress_sigex_q})
and the coarse-graining approximation are naturally expected for $q$
of order unity and larger {\em if} a grid spacing
$\agrid$ much smaller than the typical interaction range between particles is used.
%


\begin{thebibliography}{81}
\bibitem[{foo({\natexlab{b}})}]{foot_t_tau}
\bibinfo{note}{Discrete time is denoted by $t$, continuous time by $\tau$.}
\bibitem[{foo({\natexlab{c}})}]{foot_multipolar}
\bibinfo{note}{See, e.g., the wikipedia entries on quadrupoles
  and general multipolar expansions as used, say, in electrostatics. For the
  planar harmonic basis functions $\cos(p \theta)$ or $\sin(p \theta)$ relevant
  for isotropic systems in $d=2$ dimensions a monopole corresponds to $p=0$, a
  dipole to $p=1$, a quadrupole to $p=2$ and an octupole to $p=4$. We remind
  that for isotropic systems in $d=3$ both the correlation function fields and
  the associated response fields must be expanded instead in terms of spherical
  harmonic functions. Multipole expansions are related to the underlying
  isotropic symmetry and even though source terms often may not share this
  symmetry, one can expand quite generally the response field in terms of
  irreducible representations (basis functions) of the symmetry group.}
\bibitem[{foo({\natexlab{d}})}]{foot_tinf_convention}
\bibinfo{note}{In agreement with the discussion in
  Sec.~\ref{e2E_viscoelast}, ``$\tau \to \infty$" should be read ``$\tau$ is
  much larger than $\tau_1$ but many orders of magnitude smaller than
  $\tau_2$".}
\bibitem[{foo({\natexlab{e}})}]{foot_headconductivity}
\bibinfo{note}{It is assumed that temperature variations can be neglected. This
  assumption is valid by construction for our MC simulations. The isothermicity
  condition remains also true if the head conductivity is very fast compared to
  all other relaxation processes.}
\bibitem[{foo({\natexlab{f}})}]{foot_ANS_comment}
\bibinfo{note}{The generic $1/r^2$-law (cf. Fig.~\ref{fig_Cabcd_rproj}) for
  two-dimensional, isotropic and achiral systems was recently also revealed in
  the elastic-like response of a liquid probed by local strain responses to a
  localized plastic event (particle rearrangements) in a polydisperse model
  system \cite{Reichman21b}.}
\end{thebibliography}

\end{document}